\documentclass[reprint, aps, prd, showpacs,onecolumn]{revtex4-1}
\usepackage[utf8x]{inputenc}
\usepackage{amsmath, amssymb}

\allowdisplaybreaks

\begin{document}

\title{Unique Fock quantization of scalar cosmological perturbations}
\date{\today}
\author{Mikel Fern\'andez-M\'endez}
\email{m.fernandez.m@csic.es}
\author{Guillermo A. Mena Marug\'an}
\email{mena@iem.cfmac.csic.es}
\author{Javier Olmedo}
\email{olmedo@iem.cfmac.csic.es}
\affiliation{Instituto de Estructura de la Materia, IEM-CSIC, Serrano 121, 28006 Madrid, Spain}
\author{José M. Velhinho}
\email{jvelhi@ubi.pt}
\affiliation{Departamento de F\'{\i}sica, Faculdade de Ci\^encias, Universidade
da Beira Interior, R. Marqu\^es D'\'Avila e Bolama,
6201-001 Covilh\~a, Portugal.}


\begin{abstract}
We investigate the ambiguities in the Fock quantization of the scalar
perturbations of a Friedmann-Lema\^{i}tre-Robertson-Walker model with a massive
scalar field as matter content. We consider the case of compact spatial sections
(thus avoiding infrared divergences), with the topology of a three-sphere. After
expanding the perturbations in series of eigenfunctions of the Laplace-Beltrami
operator, the Hamiltonian of the system is written up to quadratic order in
them. We fix the gauge of the local degrees of freedom in two different ways,
reaching in both cases the same qualitative results. A canonical transformation,
which includes the scaling of the matter field perturbations by the scale factor
of the geometry, is performed in order to arrive at a convenient formulation of
the system. We then study the quantization of these perturbations in the
classical background determined by the homogeneous variables. Based on previous
work, we introduce a Fock representation for the perturbations in which: (a) the
complex structure is invariant under the isometries of the spatial sections and
(b) the field dynamics is implemented as a unitary operator. These two
properties select not only a unique unitary equivalence class of
representations, but also a preferred field description, picking up a canonical
pair of field variables among all those that can be obtained by means of  a
time-dependent scaling of the matter field (completed into a linear canonical
transformation). Finally, we present an equivalent quantization constructed in
terms of gauge-invariant quantities. We prove that this quantization can be
attained by a mode-by-mode time-dependent linear canonical transformation which
admits a unitary implementation, so that it is also uniquely determined.
\end{abstract}

\maketitle

\section{Introduction}

The theory of cosmological perturbations plays a major role in the explanation
of structure formation in the universe, describing the growth of originally
small perturbations in an otherwise homogeneous and isotropic background up to
the nonlinear regime. Lifschitz~\cite{Lifschitz46} was the first to study the
perturbations of the metric tensor in a Friedmann-Lema\^{i}tre-Robertson-Walker
(FLRW) model. However, it was soon realized that the interpretation of this kind
of perturbations was obscured by the freedom to perform gauge transformations,
i.e., infinitesimal coordinate transformations that keep fixed the background
spacetime. Two different ways of avoiding this problem were proposed. On the one
hand, Hawking~\cite{Hawking66} developed the first attempts to construct a fully
covariant, gauge-independent theory, which was completed by Olson for the case
of an isentropic perfect fluid with vanishing spatial curvature~\cite{Olson76}.
These works served as starting point for the covariant formulation of Ellis and
Bruni~\cite{EB89}. On the other hand, Bardeen~\cite{Bardeen80} showed how to
combine the metric and matter perturbations so as to obtain gauge-invariant
quantities. This latter approach has been extensively studied~\cite{MFB92}.

Despite the fact that perturbation theory in cosmology accounts for the growth
of the inhomogeneities that should eventually produce the large-scale structure
of the universe, by itself it does not clarify the origin of those
perturbations. At this point, additional hypotheses must be accepted. The
paradigm of cosmological inflation~\cite{LL00} provides a mechanism for
generating inhomogeneities and, moreover, solves certain problems of the
standard model of hot big-bang cosmology, such as the horizon and flatness
problems~\cite{Guth81}. According to inflation theory, the universe underwent a
period of very rapid expansion in its early history, which is generally
attributed to the presence of a hypothetical scalar field: the inflaton. The
quantum fluctuations of this field give rise to the primordial perturbations.
Among the many models of inflation that have been proposed, one of the most
popular is chaotic inflation, put forward by Linde~\cite{Linde83}. In this
model, the inflaton, subject to a potential---which can be as simple as a
quadratic one, owing to its mass---, is chaotically distributed before
inflation. The patches of the universe where the field has large values expand
exponentially, becoming nearly homogeneous in the process. One of these regions
would encompass the entire observable universe.

In order to satisfactorily understand structure formation in the universe,
therefore, a quantum or at least semiclassical description of inflation is
needed. In the absence of a complete quantum theory of gravity, one is forced to
adopt the latter approach and try to quantize the inhomogeneities in a
classical, homogeneous and isotropic background. However, this approximation
suffers from a series of problems, like for instance the generic ambiguities
that plague Quantum Field Theory in Curved Spacetime (QFTCS). These ambiguities
appear even in the case of linear fields, for which the standard procedure of
Fock quantization is available. To begin with, it is far from clear which are
the variables that one should regard as basic ones. Indeed, inasmuch as the
dynamics of the background is treated in a totally different way compared to
those of the field which is being quantized, a splitting between them must be
defined. For instance, a scaling of the variables by a time-dependent background
function (such as the scale factor in an FLRW universe) certainly modifies their
dynamics, while preserving the linearity of the system. When there is a finite
number of degrees of freedom, the classical transformations of this type (linear
but time dependent) are generally implemented as unitary transformations in the
quantum theory. Hence, the quantization of the original and of the scaled
variables are unitarily equivalent, and thus they lead to the same physical
predictions. By contrast, that is not necessarily the case in a field theory.
Moreover, there is another, more standard source of ambiguity in the
quantization of linear fields, inherent to QFTCS. It is the election of a Fock
representation of the Canonical Commutation Relations (CCR's). Once again, this
problem does not appear in systems with finite degrees of freedom, where the
Stone-von Neumann theorem~\cite{Simon72} guarantees the uniqueness (up to
unitary equivalence) of the (strongly continuous, irreducible, and unitary)
representation of the CCR's in their Weyl form. But, when dealing with an
infinite number of degrees of freedom, infinitely many inequivalent
representations are possible. This ambiguity is encoded in the choice of a
complex structure on the phase space~$\Gamma$~\cite{Wald94,CCQ04}. A complex
structure~$J$ is a linear map whose square equals minus the identity (i.e.,
$J^2=-\mathbb I$). We also require its compatibility with the symplectic
structure~$\bar{\Omega}$, in the sense that it must be a symplectomorphism and
the bilinear map $\bar{\Omega}(J\cdot,\cdot)$ must be positive-definite. In this
way, $[\bar{\Omega}(J\cdot,\cdot)-i\bar{\Omega}(\cdot,\cdot)]/2$ provides an
inner product, which endows $\Gamma$ with the structure of a complex Hilbert
space---the one-particle Hilbert space. Equivalently, one can see that a complex
structure $J$ decomposes the complexification of $\Gamma$ into two eigenspaces
of eigenvalues~$\pm i$, thus generalizing the concepts of positive and negative
frequency available in stationary scenarios. With all these ingredients, it is
possible to define the full Hilbert space of the theory, a vacuum state in it,
and finally, a representation of the CCR's.

In spite of these ambiguities, there exist criteria to select a preferred
quantization in some specific situations. A clear example is Minkowski
spacetime, where Poincar\'e symmetry permits one to select a complex structure
for a given Klein-Gordon field. In systems with a lower degree of symmetry,
other kind of conditions are required, such as energy considerations in the case
of stationary spacetimes, where a time-like Killing vector is
available~\cite{AM80}, or the existence of an invariant Gaussian solution to a
regulated Schr\"{o}dinger equation in the specific case of a de Sitter
background~\cite{jackie}. Furthermore, in the absence of stationarity, it has
been proven recently that the unitary implementation of the field dynamics is a
useful criterion to select a unique Fock representation of the CCR's (combined
with the invariance of the vacuum under any spatial symmetry remaining in the
field equations). For instance, in the Gowdy models (inhomogeneous cosmological
models with compact spatial sections and two space-like Killing
vectors~\cite{Gowdy74}), after a symmetry reduction of the system, the
inhomogeneities can be treated in a formally identical manner as a scalar field
with time-dependent mass in an auxiliary spacetime of lower dimension. The
spatial sections of the auxiliary spacetime can be circles or two-spheres,
depending on the topology of the spatial sections of the Gowdy model into
consideration. In both cases, the complex structure that would be natural if the
field were massless allows for a unitary quantum implementation of the field
dynamics and, besides, shares the symmetry of the spatial sections of the
auxiliary spacetime (which the vacuum state inherits). In fact, these two
properties select a unique unitary equivalence class of Fock representations:
all the representations with these attributes are unitarily
equivalent~\cite{CCMV06,CCMV,CMV08}. Furthermore, if we change the field
description by performing a linear canonical transformation that scales the
field by a time-dependent factor, unitarity is lost: there is no Fock
representation for the new canonical pair of field variables in which dynamics
is represented in terms of a unitary operator~\cite{CMV07}. These results also
apply in the case of a scalar field with time-dependent mass on the
three-sphere~\cite{CMV10,CMOV11}, that is to say, in a spacetime whose spatial
sections are three-dimensional spheres, a situation which is especially relevant
for our discussion in this article.

The aim of the present work is to discuss in detail the application of the
requirements of unitary evolution and symmetric vacuum state in order to remove
the ambiguities in the Fock quantization of scalar fields in a context of
particular physical interest: the scalar perturbations in an FLRW model with a
massive scalar as matter content and spatial sections with the topology of a
three-sphere. In Sec.~\ref{sec:classical}, we analyze the system classically.
Its Hamiltonian is derived in Sec.~\ref{subsec:Hamiltonian}, briefly reviewing
the work of Halliwell and Hawking~\cite{HH85}. In doing this, we expand the
perturbations in the modes of the Laplace-Beltrami operator and conduct a
perturbative analysis on them. Then, in Sec.~\ref{subsec:gauge}, we fix the
gauge of the local degrees of freedom and perform a canonical transformation of
the remaining variables so as to reach an adequate formulation of the system,
especially convenient for the rest of the analysis. In particular, we scale the
perturbation of the matter field by the FLRW scale factor. We conclude the
classical treatment writing the dynamical equations of the new canonical pair of
field variables.

We study the Fock quantization of the scalar field perturbation in the classical
background determined by the homogeneous variables in Sec.~\ref{sec:quantum}. In
Sec.~\ref{subsec:unitary}, we introduce a Fock representation that fulfills the
requirements of symmetry invariance and unitary dynamics. Any other
representation with the same properties is shown to be equivalent to it in
Sec.~\ref{subsec:equivalence}. The rest of Sec. III is devoted to the analysis
of different descriptions of the system, with other elections of canonical
variables. First, in Sec.~\ref{subsec:uniqueness}, we consider a time-dependent
scaling of the perturbation as a whole, completed into a linear canonical
transformation, and prove that the new dynamics cannot be implemented unitarily.
More general mode-by-mode transformations are then taken into account in Sec.
\ref{subsec:another}, where the system is reformulated so that a natural
quantization with unitary evolution is possible. This new quantization is
unitarily equivalent to the one previously introduced, and the Hamiltonian is
such that no cross term between the field configuration and the field momentum
appears.

In Sec.~\ref{sec:Bardeen}, we define a set of gauge-invariant quantities and
write their expressions in the chosen gauge. The dynamics of these gauge
invariants is seen to be equivalent to that of a Klein-Gordon field with
time-dependent mass. We show that any Fock representation which admits a unitary
implementation of this dynamics is unitarily equivalent to the one we already
had. Finally, a different gauge choice is explored in the Appendix.

\section{The classical model}\label{sec:classical}

\subsection{Hamiltonian of the system}\label{subsec:Hamiltonian}

We assume a globally hyperbolic spacetime and start with the ADM form of the
metric \cite{mtw}. We use the standard notation $N$ for the lapse function,
$N^i$ for the shift vector, and $h_{ij}$ for the three-metric on the spatial
sections of constant time $t$. Spatial indices are denoted with lowercase Latin
letters from the middle of the alphabet, and are lowered and raised with the
metric $h_{ij}$ and its inverse, respectively. This decomposition allows one to
express the Einstein-Hilbert action for General Relativity as the time integral
of the Lagrangian~\cite{Wald84}
\begin{equation}\label{eq:L_g}
L_g = \frac1{16\pi G}\int\!d^3\!x\,N\sqrt h\left[K_{ij}K^{ij}-\left(K^i_{\phantom{i}i}\right)^2+{}^3R\right],
\end{equation}
where $h$ is the determinant of the three-metric, $K_{ij}$ is the extrinsic curvature of the spatial sections, ${}^3R$ is their scalar curvature, and $G$ is the Newton constant.

The total Lagrangian of the system is the sum of $L_g$ and the contribution of the matter content: a scalar field $\Phi$ of mass $\tilde{m}$. The latter can be expressed in terms of the ADM variables as \cite{HH85}
\begin{align}
L_m &= \frac12\int\!d^3x \frac{\sqrt h }{N} \left[ \bigg( \frac{d\Phi}{dt} \bigg)^{2}- 2 N^i \partial_i\Phi \frac{d\Phi}{dt}  - (N^2 h^{ij}- N^iN^j )\partial_i\Phi \partial_j\Phi - N^2 \tilde{m}^2 \Phi^2 \right], \label{eq:L_m}
\end{align}
where $\partial_i$ is the flat derivative with respect to the spatial coordinate $x^i$.

In the FLRW model, $N$ depends only on time and $N^i$ vanishes, whereas the spatial sections have constant curvature. This is no longer the case if inhomogeneities are present. The description of such inhomogeneities in an FLRW background is greatly simplified by the introduction of a convenient basis of modes on the spatial sections. For the three-sphere~$S^3$, a natural basis of scalar functions (complete in the space of square-integrable functions with respect to the volume element provided by the three-metric) is given by the eigenfunctions of the Laplace-Beltrami operator, namely, the hyperspherical harmonics~\cite{HH85,GS78,jantzen}. The corresponding eigenvalue equation can be written in the form
\begin{equation}
\Omega^{ij}\nabla_i\nabla_j\tilde{Q}^n_{lm} = -\omega_n^2\tilde{Q}^n_{lm},
\end{equation}
where the symbol~$\nabla$ indicates covariant derivatives with respect to the round metric on $S^3$, which we denote by $\Omega_{ij}$. Note that its inverse is represented here by $\Omega^{ij}$, and that in FLRW we have $h_{ij}=a^2(t) \Omega_{ij}$, with $a(t)$ being the scale factor of the spatial geometry.

The eigenvalues satisfy the relation $\omega_n^2 = n^2-1$, where $n\in \mathbb{N}^+$ can be any positive integer. The corresponding eigenspaces have a degeneracy equal to $n^2$, since the labels $l$ and $m$ can take all integer values in the ranges $0\leq l \leq n-1$ and $-l\leq m \leq l $. For convenience, we adopt a basis of \emph{real} hyperspherical harmonics, because we are dealing with a real scalar field (if one started with complex harmonics, real ones can be straightforwardly obtained by taking the real and imaginary parts of the basis functions). Moreover, the hyperspherical harmonics are normalized so that they are orthonormal, in the sense that
\begin{equation}
\int\!d^3\!x\,\sqrt{\Omega}\,\tilde{Q}^n_{lm}\tilde{Q}^{n'}_{l'm'} = \delta^{nn'}\delta_{ll'}\delta_{mm'}.
\end{equation}
In the following, the labels $n$, $l$, and $m$ will be collectively denoted by $\boldsymbol n$, except if it is more convenient to show them explicitly.

Vector and tensor harmonics can be obtained from the scalar ones by means of covariant derivatives:
\begin{subequations}
\begin{align}
(\tilde{P}_i)^{\boldsymbol n} = \frac1{\omega_n^2}\nabla_i \tilde{Q}^{\boldsymbol n}, \quad
(\tilde{P}_{ij})^{\boldsymbol n} = \nabla_j(\tilde{P}_i)^{\boldsymbol n}+\frac13\Omega_{ij}\tilde{Q}^{\boldsymbol n}, \quad  n\in\{2,3,\ldots\}.
\end{align}
\end{subequations}
These harmonics are not complete in the respective spaces of vector and tensor fields on $S^3$, as there exist other vector and tensor harmonics which cannot be obtained in this way. Nevertheless, we will consider only inhomogeneities constructed from scalar perturbations (and their derivatives). A complete analysis shows that the dynamics of such perturbations is completely decoupled from the genuinely vectorial and tensorial ones~\cite{HH85}.

The variables of the model can be expanded in series of hyperspherical harmonics, the $n=1$ term corresponding to the homogeneous variables of the FLRW model. We can make this explicit by writing
\begin{gather}\label{eqs:variables}
h_{ij} = \sigma^2e^{2\alpha}\Omega_{ij}\left(1+2\sqrt2\,\pi\epsilon\sum_{\boldsymbol n}a_{\boldsymbol n}\tilde{Q}^{\boldsymbol n}\right) +\sigma^2e^{2\alpha}6\sqrt2\,\pi\epsilon\sum_{\boldsymbol n}b_{\boldsymbol n}(\tilde{P}_{ij})^{\boldsymbol n}, \\ \nonumber
N = \sigma N_0\left(1+\sqrt2\,\pi\epsilon\sum_{\boldsymbol n}g_{\boldsymbol n}\tilde{Q}^{\boldsymbol n}\right), \quad N_i = \sigma^2e^\alpha\sqrt2\,\pi\epsilon\sum_{\boldsymbol n}k_{\boldsymbol n}(\tilde{P}_i)^{\boldsymbol n}, \quad
\Phi = \frac1\sigma\left(\frac1{\sqrt2\,\pi}\varphi+\epsilon\sum_{\boldsymbol n}f_{\boldsymbol n}\tilde{Q}^{\boldsymbol n}\right).
\end{gather}
All these sums start in $n=2$, except the metric contribution with coefficients~$b_{\boldsymbol n}$, which begins in $n=3$ because the corresponding tensor harmonic with $n=2$ turns out to be proportional to $\Omega_{ij}$. The factor $\sigma=\sqrt{2G/3\pi}$ has been introduced for convenience. The ``background'' functions $N_0$, $\alpha$, and $\varphi$ parameterize the homogeneous solutions, whereas the coefficients $a_{\boldsymbol n}$, $b_{\boldsymbol n}$, $g_{\boldsymbol n}$, $k_{\boldsymbol n}$, and $f_{\boldsymbol n}$ describe the inhomogeneities. They are all functions of time only, since the hyperspherical harmonics absorb all the spatial dependence (apart from that of the components of the metric~$\Omega_{ij}$). We assume that the inhomogeneities are sufficiently small so as to justify a perturbative analysis up to second order in the parameter $\epsilon$ (namely, up to second order in the inhomogeneities). Note that if all the perturbations are neglected, the FLRW model is recovered, with $\sigma N_0$ being the homogeneous lapse function, $\sigma e^\alpha$ the scale factor, and $({\sqrt2\,\pi\sigma})^{-1}\varphi$ being a homogeneous scalar field of mass~$\tilde{m}=m/\sigma$.

Introducing expressions~\eqref{eqs:variables} in Eqs.~\eqref{eq:L_g} and~\eqref{eq:L_m}, one can obtain the Lagrangian of the system within the perturbative framework~\cite{HH85}. A Legendre transform with respect to the time derivatives of the variables $\alpha$, $\varphi$, $a_{\boldsymbol n}$, $b_{\boldsymbol n}$, and $f_{\boldsymbol n}$ leads to a Hamiltonian of the form
\begin{align}
H &= N_0\left[H_{|0}+ \epsilon^2\sum_{\boldsymbol n}\left(H^{\boldsymbol n}_{|2}+ g_{\boldsymbol n}H^{\boldsymbol n}_{|1}\right)\right] +\epsilon^2 \sum_{\boldsymbol n}k_{\boldsymbol n}H^{\boldsymbol n}_{\_1}+O(\epsilon^3),
\end{align}
which is a linear combination of constraints, with $N_0$, $N_0g_{\boldsymbol n}$, and $k_{\boldsymbol n}$ acting as Lagrange multipliers. The subscripts of the quantities appearing in the last expression indicate their association with the lapse function~($|$) or the shift vector~($\_$), as well as their perturbative order in terms of the inhomogeneous modes (namely, their degree as polynomials of the coefficients of the inhomogeneities).

Calling generically $\pi_q$ the momentum conjugate to any variable $q$ \cite{note1}, the zeroth-order Hamiltonian is written as
\begin{equation}\label{eq:H_0}
H_{|0} = \frac12e^{-3\alpha}\left(-\pi_\alpha^2+\pi_\varphi^2+e^{6\alpha}m^2\varphi^2-e^{4\alpha}\right),
\end{equation}
while the higher-order terms have the explicit form
\begin{widetext}
\begin{align}
H^{\boldsymbol n}_{|2} &= \frac{1}{2}e^{-3\alpha}\Bigg[\left(\frac12a_{\boldsymbol n}^2+10\frac{n^2-4}{n^2-1}b_{\boldsymbol n}^2\right)\pi_\alpha^2+ \left(\frac{15}2a_{\boldsymbol n}^2+6\frac{n^2-4}{n^2-1}b_{\boldsymbol n}^2\right)\pi_\varphi^2 \nonumber\\*
&\phantom{=\frac12e^{-3\alpha}\Bigg[}-\pi_{a_{\boldsymbol n}}^2+ \frac{n^2-1}{n^2-4}\pi_{b_{\boldsymbol n}}^2+\left(2a_{\boldsymbol n}\pi_{a_{\boldsymbol n}}+8b_{\boldsymbol n}\pi_{b_{\boldsymbol n}}\right)\pi_\alpha+\pi_{f_{\boldsymbol n}}^2-6a_{\boldsymbol n}\pi_\varphi\pi_{f_{\boldsymbol n}} \nonumber\\*
&\phantom{=\frac12e^{-3\alpha}\Bigg[}  -e^{4\alpha}\Bigg\{\frac13\left(n^2-\tfrac52\right)a_{\boldsymbol n}^2+\frac13\left(n^2-7\right)\frac{n^2-4}{n^2-1}b_{\boldsymbol n}^2+\frac23\left(n^2-4\right)a_{\boldsymbol n}b_{\boldsymbol n}-\left(n^2-1\right)f_{\boldsymbol n}^2\Bigg\} \nonumber\\*
&\phantom{=\frac12e^{-3\alpha}\Bigg[} +e^{6\alpha}m^2\left(\frac32\varphi^2a_{\boldsymbol n}^2-6\frac{n^2-4}{n^2-1}\varphi^2b_{\boldsymbol n}^2+6\varphi a_{\boldsymbol n}f_{\boldsymbol n}+f_{\boldsymbol n}^2\right)\Bigg], \\
H^{\boldsymbol n}_{|1} &= \frac{1}{2}e^{-3\alpha}\Bigg[-a_{\boldsymbol n}\pi_\alpha^2-3a_{\boldsymbol n}\pi_\varphi^2-2\pi_\alpha\pi_{a_{\boldsymbol n}}+2\pi_\varphi\pi_{f_{\boldsymbol n}}-\frac23e^{4\alpha}\left[\left(n^2+\tfrac12\right)a_{\boldsymbol n}+\left(n^2-4\right)b_{\boldsymbol n}\right] \nonumber\\*
&\phantom{= \frac{1}{2}e^{-3\alpha}\Bigg[} +e^{6\alpha}m^2\varphi\left(3\varphi a_{\boldsymbol n}+2f_{\boldsymbol n}\right)\Bigg], \\
H^{\boldsymbol n}_{\_1} &= \frac{1}{3}e^{-\alpha}\Bigg[\left(a_{\boldsymbol n}+4\frac{n^2-4}{n^2-1}b_{\boldsymbol n}\right)\pi_\alpha+3f_{\boldsymbol n}\pi_\varphi-\pi_{a_{\boldsymbol n}}+\pi_{b_{\boldsymbol n}}\Bigg].
\end{align}
\end{widetext}

\subsection{Gauge fixing and reformulation of the system}\label{subsec:gauge}

The presence of constraints indicates that the system includes unphysical degrees of freedom. Here, we partially fix this gauge freedom by adopting the longitudinal gauge. A different gauge choice is addressed in the Appendix, to show that the results are not sensitive to the particular gauge adopted.

More specifically, we want to remove the freedom associated with the Lagrange multipliers $g_{\boldsymbol n}$ and $k_{\boldsymbol n}$ (corresponding to the constraints $H_{|1}^{\boldsymbol n}=0$ and $H_{\_1}^{\boldsymbol n}=0$, respectively), setting  $N_i=0$ and $h_{ij}\propto\Omega_{ij}$. To this end, we impose
\begin{gather}\label{eqs:gauge1}
b_{\boldsymbol n} = 0, \quad C_{\boldsymbol n} \equiv -a_{\boldsymbol n}\pi_\alpha-3f_{\boldsymbol n}\pi_\varphi+\pi_{a_{\boldsymbol n}} = 0.
\end{gather}
These are acceptable gauge-fixing conditions as far as $n>2$, since in this case their Poisson brackets with $H_{\_1}^{\boldsymbol n}$ and $H_{|1}^{\boldsymbol n}$ satisfy
\begin{equation}
\epsilon^4 \det
\begin{pmatrix}
\{b_{\boldsymbol n}, H_{\_1}^{\boldsymbol n}\} & \{C_{\boldsymbol n}, H_{\_1}^{\boldsymbol n}\} \\
\{b_{\boldsymbol n}, H_{|1}^{\boldsymbol n}\} & \{C_{\boldsymbol n}, H_{|1}^{\boldsymbol n}\}
\end{pmatrix}
\approx \frac19(n^2-4) > 0.
\end{equation}
The symbol $\approx$ indicates that the identity holds when one makes use of the constraints and the gauge-fixing conditions.

Eqs.~\eqref{eqs:gauge1}, together with the constraints, allow one to write $a_{\boldsymbol n}$ and $\pi_{b_{\boldsymbol n}}$ in terms of $f_{\boldsymbol n}$, $\pi_{f_{\boldsymbol n}}$, and the homogeneous variables. In particular,
\begin{equation}
a_{\boldsymbol n} \approx 3\frac{(e^{6\alpha}m^2\varphi-3\pi_\alpha\pi_\varphi)f_{\boldsymbol n}+\pi_\varphi\pi _{f_{\boldsymbol n}}}{9\pi_\varphi^2+(n^2-4)e^{4\alpha}}+O(\epsilon^2),
\end{equation}
whereas $\pi_{b_n}=0$. Besides, consistency of the gauge-fixing conditions with the dynamical evolution requires the imposition of the following additional restrictions:
\begin{align}
0 = \{b_{\boldsymbol n},H\} & \approx \frac13e^{-\alpha}k_{\boldsymbol n}+O(\epsilon), \qquad
0 = \{C_{\boldsymbol n},H\}  \approx \frac13(n^2-4)N_0e^\alpha(a_{\boldsymbol n}+g_{\boldsymbol n})+O(\epsilon).
\end{align}
Therefore, we get $k_{\boldsymbol n}=0$ and $g_{\boldsymbol n}=-a_{\boldsymbol n}$ up to higher-order perturbative corrections.

In the case $n=2$, conditions \eqref{eqs:gauge1} are no longer applicable since, by construction, the coefficients $b_{\boldsymbol n}$ exist only for $n>2$. Actually, for the second mode, one can always fix the gauge so that $a_{2lm}=f_{2lm}=0$. This leaves no remnant physical degree of freedom for this mode.

After this gauge fixation, the spacetime metric reads
\begin{align}\label{eq:metric}
ds^2 = -\sigma^2N_0^2\left(1-2\sqrt2\,\pi\epsilon\sum_{\boldsymbol n}a_{\boldsymbol n}\tilde{Q}^{\boldsymbol n}\right)dt^2+\sigma^2e^{2\bar\alpha}\Omega_{ij}\left(1+2\sqrt2\,\pi\epsilon\sum_{\boldsymbol n}a_{\boldsymbol n}\tilde{Q}^{\boldsymbol n}\right)dx^idx^j+O(\epsilon^2).
\end{align}

In the reduction of the system, it is convenient to introduce new coordinates on phase space so as to reach a canonical set with respect to the new, reduced symplectic structure. Namely, we keep $\pi_\alpha$, $\pi_\varphi$, and $f_{\boldsymbol n}$, while their conjugate variables \cite{note1} become
\begin{align}\label{eqs:tilde}
\tilde\alpha = \alpha+\frac{\epsilon^2}2\sum_{\boldsymbol n}a_{\boldsymbol n}^2, \qquad \tilde\varphi = \varphi +3\epsilon^2\sum_{\boldsymbol n}a_{\boldsymbol n}f_{\boldsymbol n}, \qquad
\tilde\pi_{f_{\boldsymbol n}} = \pi_{f_{\boldsymbol n}}-3a_{\boldsymbol n}\pi_\varphi.
\end{align}
The reduced Hamiltonian can be written as
\begin{equation}\label{eq:H}
\tilde H = N_0\left(\tilde H_{|0}+\epsilon^2\sum_{\boldsymbol n}\tilde H^{\boldsymbol n}_{|2}\right)+O(\epsilon^3).
\end{equation}
Since the new homogeneous variables differ from the old ones only by quadratic terms in the perturbations, the expression of the zeroth-order Hamiltonian is unaffected by the transformation: it is still given by the right-hand side of Eq.~\eqref{eq:H_0}, provided that the old variables are replaced with the new ones. As for the second-order Hamiltonian, it is of the form
\begin{equation}\label{eq:H_2}
\tilde H^{\boldsymbol n}_{|2} = \frac{1}2e^{-\tilde\alpha}\left(\tilde E^n_{\pi\pi}\tilde\pi_{f_{\boldsymbol n}}^2+2\tilde E^n_{f\pi}f_{\boldsymbol n}\tilde\pi_{f_{\boldsymbol n}}+\tilde E^n_{ff}f_{\boldsymbol n}^2\right),
\end{equation}
where we have introduced the notation
\begin{subequations}
\begin{align}
\tilde E^n_{\pi\pi} &= e^{-2\tilde\alpha}\left(1-\frac{3e^{-4\tilde\alpha}}{n^2-4}\pi_\varphi^2\right), \\
\tilde E^n_{f\pi} &= \frac{3}{n^2-4}e^{-6\tilde\alpha}\pi_\varphi\left(3\pi_\alpha\pi_\varphi-e^{6\tilde\alpha}m^2\tilde\varphi\right), \label{tEnfp}\\
\tilde E^n_{ff} &= e^{2\tilde\alpha}\left(n^2-1\right)+e^{4\tilde\alpha}m^2-9e^{-2\tilde\alpha}\pi_\varphi^2  -\frac{3}{n^2-4}e^{-6\tilde\alpha}\left(3\pi_\alpha\pi_\varphi-e^{6\tilde\alpha}m^2\tilde\varphi\right)^2.
\end{align}
\end{subequations}

Before concluding the classical treatment of the system, we perform a canonical transformation that reformulates the system in a especially suitable form. First of all, we scale the perturbation of the matter field by the FLRW scale factor: $f_{\boldsymbol n}\mapsto e^\alpha f_{\boldsymbol n}$. This transformation (that we will call the Mukhanov scaling; see, e.g., Ref.~\cite{mukhanov}) is frequently adopted when working in an FLRW background, as it simplifies the equations of motion. We can complete this change into a canonical transformation by introducing the inverse scaling in the conjugate field momentum. Besides, we allow also for a contribution to this new momentum that is linear in the field configuration, with the aim at removing cross terms in the Hamiltonian which would couple the configuration and momentum variables of the perturbations---at least up to subdominant terms in the large-$n$ limit. Apart from further simplifying the system, in this way we adapt its description to the most convenient form for the asymptotic analysis of the forthcoming section. These new variables, which have canonical Poisson brackets at the considered order in the perturbations, are
\begin{gather}\label{eqs:transformation}
\bar\alpha = \tilde\alpha+\frac{\epsilon^2}2\sum_{\boldsymbol n}f_{\boldsymbol n}^2, \qquad
\bar\pi_{\bar\alpha} = \pi_\alpha+\epsilon^2\sum_{\boldsymbol n}\left(-f_{\boldsymbol n}\tilde\pi_{f_{\boldsymbol n}}+\pi_\alpha f_{\boldsymbol n}^2\right), \qquad
\bar f_{\boldsymbol n} = e^{\tilde\alpha}f_{\boldsymbol n}, \qquad
\bar\pi_{\bar{f}_{\boldsymbol n}} = e^{-\tilde\alpha}\left(\tilde\pi_{f_{\boldsymbol n}}-\pi_\alpha f_{\boldsymbol n}\right),
\end{gather}
while $\tilde\varphi$ and $\pi_\varphi$ are left unchanged.

In terms of these new variables, the Hamiltonian has the same structure as in Eqs.~\eqref{eq:H} and~\eqref{eq:H_2}, with the functional form of the zeroth-order Hamiltonian still being given by Eq.~\eqref{eq:H_0}. On the other hand, the coefficients of the second-order Hamiltonian are now
\begin{subequations}\label{eqs:E^n}
\begin{align}
\bar E^n_{\pi\pi} &= 1-\frac{3}{n^2-4}e^{-4\bar\alpha}\pi_\varphi^2, \\
\bar E^n_{f\pi} &= \frac{3}{n^2-4}e^{-6\bar\alpha}\pi_\varphi\left(2\bar\pi_{\bar\alpha}\pi_\varphi-e^{6\bar\alpha}m^2\tilde\varphi\right), \\
\bar E^n_{ff} &= n^2-\frac12+\frac12 e^{2\bar\alpha}m^2(2-3\tilde\varphi^2)-\frac{1}2e^{-4\bar\alpha}\left(\bar\pi_{\bar\alpha}^2+15\pi_\varphi^2\right)-\frac{3}{n^2-4}e^{-8\bar\alpha}\left(2\bar\pi_{\bar\alpha}\pi_\varphi-e^{6\bar\alpha}m^2\tilde\varphi\right)^2.
\end{align}
\end{subequations}
Thus, as we had anticipated, we have implemented the Mukhanov scaling while retaining the behavior of the cross terms in the Hamiltonian, $\bar E^n_{f\pi}=O(n^{-2})$, again like $\tilde E^n_{f\pi}$ in Eq.~\eqref{tEnfp}. 

On the other hand, the expression of $a_{\boldsymbol n}$ as a function of these variables is
\begin{align}\label{eq:a_n}
a_{\boldsymbol n} = \frac{3}{n^2-4}&\left[e^{-5\bar\alpha}\left(e^{6\bar\alpha}m^2\tilde\varphi-2\bar\pi_{\bar\alpha}\pi_\varphi\right)\bar f_{\boldsymbol n}+e^{-3\bar\alpha}\pi_\varphi\bar\pi_{\bar f_{\boldsymbol n}}\right].
\end{align}
Taking this into account, the spacetime metric can be easily obtained via Eq.~\eqref{eq:metric}.

To end this section, we will write the dynamical equations of the system. In doing this, we will use the conformal time~$\eta$, defined by $N_0dt=e^\alpha d\eta$. Derivatives with respect to this time will be denoted by a dot. Hamilton's equations give
\begin{subequations}\label{eqs:Hamiltoneqs}
\begin{gather}
\dot{\bar\alpha} = -e^{-2\bar\alpha}\bar\pi_{\bar\alpha}+O(\epsilon^2), \qquad
\dot{\bar\pi}_{\bar\alpha} = \frac{e^{-2\bar\alpha}}2\big(-3\bar\pi_{\bar\alpha}^2+3\pi_\varphi^2-3e^{6\bar\alpha}m^2\tilde\varphi^2+e^{4\bar\alpha}\big)+O(\epsilon^2), \\
\dot{\tilde\varphi} = e^{-2\alpha}\pi_\varphi+O(\epsilon^2), \qquad
\dot\pi_\varphi = -e^{4\alpha}m^2\tilde\varphi+O(\epsilon^2), \\
\dot{\bar f}_{\boldsymbol n} = \left(\bar E^n_{\pi\pi}\bar\pi_{\bar f_{\boldsymbol n}}+\bar E^n_{f\pi}\bar f_{\boldsymbol n}\right)+O(\epsilon), \label{eq:dotf}\\
\dot{\bar\pi}_{\bar f_{\boldsymbol n}} = -\left(\bar E^n_{f\pi}\bar\pi_{\bar f_{\boldsymbol n}}+\bar E^n_{ff}\bar f_{\boldsymbol n}\right)+O(\epsilon).
\end{gather}
\end{subequations}
Using Eq.~\eqref{eq:dotf}, we can isolate $\bar\pi_{\bar f_{\boldsymbol n}}$ and write it as
\begin{equation}\label{eq:momentum}
\bar\pi_{\bar f_{\boldsymbol n}} = (1+p_n)\dot{\bar f}_{\boldsymbol n}+q_n\bar f_{\boldsymbol n},
\end{equation}
where we have neglected higher-order contributions and introduced the time-dependent coefficients
\begin{align}\label{pq_n}
p_n = \frac{3\pi_\varphi^2}{(n^2-4)e^{4\bar\alpha}-3\pi_\varphi^2}, \qquad
q_n = -3e^{-2\bar\alpha}\pi_\varphi\frac{2\bar\pi_{\bar\alpha}\pi_\varphi-e^{6\bar\alpha}m^2\tilde\varphi}{(n^2-4)e^{4\bar\alpha}-3\pi_\varphi^2}.
\end{align}
Combining the two last equations in the set~\eqref{eqs:Hamiltoneqs}, we obtain the equation of motion of $f_{\boldsymbol n}$:
\begin{align}
\left(\frac{\dot{\bar E}^n_{\pi\pi}}{\bar E^n_{\pi\pi}}\bar E^n_{f\pi}-\dot{\bar E}^n_{f\pi}-\big(\bar E^n_{f\pi}\big)^2+\bar E^n_{ff}\bar E^n_{\pi\pi}\right)\bar f_{\boldsymbol n}
-\frac{\dot{\bar E}^n_{\pi\pi}}{\bar E^n_{\pi\pi}}\dot{\bar f}_{\boldsymbol n}+\ddot{\bar f}_{\boldsymbol n} = O(\epsilon),
\end{align}
which can be written in the simpler form (neglecting again higher-order terms)
\begin{equation}\label{eq:theequation}
\ddot{\bar f}_{\boldsymbol n}+r_n\dot{\bar f}_{\boldsymbol n}+(\omega_n^2+s_n)\bar f_{\boldsymbol n}=0,
\end{equation}
where
\begin{subequations}
\begin{align}
r_n &= 6e^{-2\bar\alpha}\pi_\varphi\frac{2\bar\pi_{\bar\alpha}\pi_\varphi-e^{6\bar\alpha}m^2\tilde\varphi}{(n^2-4)e^{4\bar\alpha}-3\pi_\varphi^2}, \label{eq:r_n}\\
s_n &= \frac12+e^{2\bar\alpha}m^2-\frac{1}2e^{-4\bar\alpha}\left(\bar\pi_{\bar\alpha}^2+21\pi_\varphi^2+3e^{6\bar\alpha}m\tilde\varphi^2\right)+O\left(n^{-2}\right). \label{eq:s_n}
\end{align}
\end{subequations}

\section{Fock quantization of the perturbations}\label{sec:quantum}

Let us study now the Fock quantization of the scalar field perturbation in the background determined by the homogeneous variables, which we will describe classically (to all purposes, we will treat them as
mere functions of time). This semiclassical treatment should provide a good approximation to the behavior of the system as long as quantum effects on the background do not become important and one can disregard non-conventional quantum geometry phenomena (e.g., those of polymeric type predicated by Loop Quantum Gravity \cite{lqg,lqc}) in the inhomogeneities. Thus, we focus on the (standard) quantum fluctuations of only the local degrees of freedom, neglecting quantum corrections that could arise from the zero modes parametrized by the variables $\alpha$, $\phi$, and their momenta.

The arguments presented in the following two subsections are similar to those detailed in Ref.~\cite{CMV10} for the Fock quantization of a Klein-Gordon field with a time-dependent mass on $S^3$. Despite the differences between the two situations, the analysis can be adapted without major difficulties. In order to make this manuscript self contained, we now summarize the master lines of the proof discussed in Ref. \cite{CMV10}, and explain the modifications needed for our specific model.

\subsection{A quantization with unitary dynamics}\label{subsec:unitary}

As we have commented in the Introduction, QFTCS is plagued with ambiguities. These include the choice of a field description. Among all those that can be reached by a time-dependent scaling of the field, we will initially select the canonical pair formed by the perturbative modes~$\bar f_{\boldsymbol n}$ and their conjugate momenta~$\bar\pi_{\bar f_{\boldsymbol n}}$, although we will discuss other possible options later on. In the previous section, we derived the equation of motion~\eqref{eq:theequation} for $\bar f_{\boldsymbol n}$. In particular, note that all the modes with the same value of $n$ evolve in the same way. Besides, in the asymptotic limit $n\rightarrow\infty$, one recovers the equation for the modes of a free massless scalar field on $S^3$, at least apparently, because the function $r_n$ is negligible [see Eq.~\eqref{eq:r_n}], and the function $s_n$ [in Eq.~\eqref{eq:s_n}] is much smaller than $\omega_n^2$. On the other hand, the momentum~$\bar\pi_{\bar f_{\boldsymbol n}}$ obeys Eq.~\eqref{eq:momentum}, and hence differs from the time derivative of $\bar f_{\boldsymbol n}$ only by terms of the order of $n^{-2}$.

Even after selecting a canonical pair, there is still an infinite ambiguity in the choice of a Fock representation for the corresponding CCR's. This freedom in the construction of a Fock representation amounts to the selection of a complex structure that is compatible with the symplectic
structure. To fix this complex structure, let us first introduce the annihilation and creation variables that would be natural to adopt in the case of a free massless field:
\begin{equation}\label{eq:creation}
\begin{pmatrix}
a_{\bar f_{\boldsymbol n}} \\
a_{\bar f_{\boldsymbol n}}^*
\end{pmatrix}
= \frac1{\sqrt{2\omega_n}}
\begin{pmatrix}
\omega_n & i \\
\omega_n & -i
\end{pmatrix}
\begin{pmatrix}
\bar f_{\boldsymbol n} \\
\bar\pi_{\bar f_{\boldsymbol n}}
\end{pmatrix},
\end{equation}
with the symbol $^*$ standing for complex conjugation.
Then, we choose the complex structure $J_0$ that is diagonal in this basis of phase space variables:
\begin{equation}
J_0
\begin{pmatrix}
a_{\bar f_{\boldsymbol n}} \\
a_{\bar f_{\boldsymbol n}}^*
\end{pmatrix}
=
\begin{pmatrix}
i & 0 \\
0 & -i
\end{pmatrix}
\begin{pmatrix}
a_{\bar f_{\boldsymbol n}} \\
a_{\bar f_{\boldsymbol n}}^*
\end{pmatrix}.
\end{equation}
In the following, we will call $j_0$ the $2\times2$ matrix that appears in this last equation.

The complex structure introduced above shares the symmetry of the spatial sections, i.e., it is invariant under SO(4) rotations. It is clear that such transformations do not mix variables with different label $n$, as they commute with the Laplace-Beltrami operator. Moreover,
the configuration and momentum spaces associated with a given fixed $n$, namely the sets
\begin{equation}
\mathcal F_n = \{\bar f_{nlm};\text{ fixed }n\}, \quad  \mathcal P_n = \{\bar \pi_{\bar f_{nlm}};\text{ fixed }n\},
\end{equation}
transform with the same irreducible representation of SO(4)~\cite{CMV10}, so creation variables do not mix with annihilation variables under the kind of transformations that we are considering.

Owing to the fact that the modes are dynamically decoupled, the symplectomorphism~$\mathcal U$ that gives their time evolution is block diagonal. Thus, fixing a(n arbitrary) reference time~$\eta_0$, we can write
\begin{equation}\label{eq:classevo}
\begin{pmatrix}
a_{\bar f_{\boldsymbol n}}(\eta) \\
a_{\bar f_{\boldsymbol n}}^*(\eta)
\end{pmatrix}
= \mathcal U_n(\eta,\eta_0)
\begin{pmatrix}
a_{\bar f_{\boldsymbol n}}(\eta_0) \\
a_{\bar f_{\boldsymbol n}}^*(\eta_0)
\end{pmatrix}.
\end{equation}
The matrix~$\mathcal U_n$ depends
on $n$, but not on the degeneracy labels $l$ and $m$, because the evolution is insensitive to them. Besides, since $\mathcal U_n$ represents a real transformation, it has the general form
\begin{equation}\label{eq:U_n}
\mathcal U_n(\eta,\eta_0) =
\begin{pmatrix}
\alpha_n(\eta,\eta_0) & \beta_n(\eta,\eta_0) \\
\beta_n^*(\eta,\eta_0) & \alpha_n^*(\eta,\eta_0)
\end{pmatrix}.
\end{equation}
Furthermore, as the dynamics is a symplectomorphism, the Bogoliubov coefficients $\alpha_n$ and $\beta_n$ satisfy the relation $|\alpha_n|^2-|\beta_n|^2=1$, so that Poisson brackets remain invariant under the transformation.

In the quantum theory, annihilation and creation variables $a_{\boldsymbol n}$ and $a_{\boldsymbol n}^*$ are promoted to operators $\hat a_{\boldsymbol n}$ and $\hat a_{\boldsymbol n}^\dagger$, whose evolution is given by an analogue of Eq.~\eqref{eq:classevo}:
\begin{equation}\label{eq:quantevo}
\begin{pmatrix}
\hat a_{\bar f_{\boldsymbol n}}(\eta) \\
\hat a_{\bar f_{\boldsymbol n}}^\dagger(\eta)
\end{pmatrix}
= \mathcal U_n(\eta,\eta_0)
\begin{pmatrix}
\hat a_{\bar f_{\boldsymbol n}}(\eta_0) \\
\hat a_{\bar f_{\boldsymbol n}}^\dagger(\eta_0)
\end{pmatrix}.
\end{equation}
The complex structure~$J_0$ has the particular property that the Hilbert-space operator that implements the above evolution exists and is unitary, as we will now prove. The necessary and sufficient condition for that to happen is that
\begin{equation}
J_0-\mathcal U(\eta,\eta_0)J_0\mathcal U^{-1}(\eta,\eta_0)
\end{equation}
must be a Hilbert-Schmidt operator (for all times $\eta$) on the one-particle Hilbert space defined by $J_0$~\cite{Shale62,HR96}. This condition guarantees that the evolution of the vacuum state at $\eta_0$ is a (normalizable) state at all times, and can be rephrased as the requirement that the beta-coefficients of the dynamical transformation are square summable (in complex norm), namely:
\begin{equation}
\sum_{\boldsymbol n}|\beta_n(\eta,\eta_0)|^2 < \infty,
\end{equation}
which, recalling the $n^2$-fold degeneracy of each eigenvalue~$n$, is equivalent to demanding the square sum\-mability of the sequence $\{n\beta_n\}_n$. Consequently, the unitary implementation of the dynamics relies only on the asymptotic behavior of the Bogoliubov coefficients $\beta_n$ for large $n$, which can be deduced from the equation of motion~\eqref{eq:theequation}. In order to study the asymptotic limit of the solutions, we write them in the form
\begin{equation}\label{eq:asymptotic}
\bar f_{\boldsymbol n}(\eta) = A_{\boldsymbol n}e^{\omega_n\Theta_n(\eta)}+A_{\boldsymbol n}^*e^{\omega_n\Theta_n^*(\eta)},
\end{equation}
where $A_{\boldsymbol n}$ is a complex parameter that is determined by the initial conditions on $f_{\boldsymbol n}$ (i.e., the initial values of the mode and of its time derivative). Actually, this relation depends on the initial values of the complex function~$\Theta_n$. One can
suitably fix these, without loss of generality, so that $\Theta_n(\eta_0)=0$ and $\dot\Theta_n(\eta_0)=-i$ \cite{CMSV09}. Then, one gets
\begin{equation}\label{eq:A_n}
A_{\boldsymbol n} = \frac1{2\omega_n}[\omega_n\bar f_{\boldsymbol n}(\eta_0)+i\dot{\bar f}_{\boldsymbol n}(\eta_0)].
\end{equation}
In the free massless case (regime which the system is expected to approach as $n\rightarrow\infty$), one would have $\dot\Theta_n(\eta)=-i$ $\forall\eta$. It is convenient to reexpress $\Theta_n$ so that this contribution is extracted:
\begin{equation}\label{eq:Theta_n}
\Theta_n(\eta) = -i(\eta-\eta_0)+\int_{\eta_0}^\eta\frac{W_n(\bar\eta)}{\omega_n}d\bar\eta.
\end{equation}
The complex function~$W_n$ that we have introduced satisfies $W_n(\eta_0)=0$. Eq.~\eqref{eq:theequation} now transforms into the Ricatti equation
\begin{equation}\label{eq:W_n}
\dot W_n = i\omega_nr_n-s_n+(2i\omega_n-r_n)W_n-W_n^2.
\end{equation}
A careful but straightforward asymptotic analysis of this equation in the large-$n$ limit, similar to that presented in Ref.~\cite{CMV10}, leads one to conclude that its solutions are at most of the order of $n^{-1}$. Neglecting explicitly subdominant terms in the right-hand side of Eq.~\eqref{eq:W_n}, we arrive at
\begin{equation}
\dot W_n = -s+2i\omega_nW_n+O(n^{-1}),
\end{equation}
where $s=\lim_{n\rightarrow\infty}s_n$ denotes the leading contribution to the function $s_n$, up to corrections of order $O(n^{-2})$. Note that $s$ is well defined [see Eq.~\eqref{eq:s_n}]. With the initial condition $W_n(\eta_0)=0$, we obtain
\begin{align}\label{eq:asymptoticW_n}
W_n(\eta) &= \frac1{2i\omega_n}\left[s(\eta)-s(\eta_0)e^{2i\omega_n(\eta-\eta_0)}\right] 
-\frac{e ^{2i\omega_n\eta}}{2i\omega_n}\int_{\eta_0}^\eta \dot s(\bar\eta)e^{-2i\omega_n\bar\eta}d\bar\eta+O(n^{-2}).
\end{align}
Assuming that the function $s$ is differentiable and its derivative $\dot s$ is integrable in every closed subinterval of the (possibly unbounded) interval of definition of the time $\eta$, we then see that the solution is indeed of the order of $n^{-1}$.

On the other hand, a simple calculation shows that the coefficients $\alpha_n$ and $\beta_n$ have the following expressions in terms of the functions $W_n$ and $\Theta_n$:
\begin{subequations}
\begin{align}
\alpha_n(\eta,\eta_0) &=  \frac{e^{\omega_n\Theta_n(\eta)}}{1+p_n(\eta_0)}\bigg\{\big[1+D_n(\eta)\big]\big[1+B_n(\eta_0)\big]
-D_n^*(\eta)B_n^*(\eta_0)\bigg\}, \label{eq:alpha_n}\\
\beta_n(\eta,\eta_0) &=  \frac{e^{\omega_n\Theta_n(\eta)}}{1+p_n(\eta_0)}\bigg\{\big[1+D_n(\eta)\big]B_n(\eta_0)
-D_n^*(\eta)\big[1+B_n^*(\eta_0)\big]\bigg\},
\end{align}
\end{subequations}
where
\begin{align}
B_n = \frac12\left(p_n-i\frac{q_n}{\omega_n}\right), \qquad D_n = B_n^*+\frac i2(1+p_n)\frac{W_n}{\omega_n},
\end{align}
and $p_n$ and $q_n$ are given in Eqs.~\eqref{pq_n}.

Using these relations, it can be easily checked that the coefficients $\beta_n$ are of the asymptotic order of $n^{-2}$. As a result, the sequence $\{n\beta_n\}_n$ turns out to be square summable, and the dynamics is unitarily implementable. In comparison with the situation discussed in Ref.~\cite{CMV10}, the asymptotic corrections to the equations of motion provided by the functions $p_n$, $q_n$, $r_n$, or the $O(n^{-2})$ contributions to $s_n$, are subdominant enough so as to preserve the unitary implementability of the dynamics.

\subsection{Equivalence of the invariant representations with unitary dynamics}\label{subsec:equivalence}

The quantum representation we have presented is not the only SO(4)-invariant one in which the dynamics is implemented by a unitary operator. Nonetheless, we can prove that any other representation with an SO(4)-invariant vacuum state and unitary dynamics must be unitarily equivalent to the one given above.

In the following, we will call invariant the complex structures that commute with the action of the group of SO(4) transformations. Actually, an invariant complex structure cannot mix modes differing in any of the labels $n$, $l$, or $m$. Firstly, since the eigenspaces of the Laplace-Beltrami operator on $S^3$ are actually irreducible representations of the SO(4) group, such eigenspaces must be preserved by any invariant complex structure~$J$. In consequence, $J$ has the general form $J=\bigoplus_nJ_n$, where each complex structure $J_n$ acts exclusively on the subspace $\mathcal F_n\oplus\mathcal P_n$ of the phase space. Furthermore, $J_n$ consists of four components, namely, the maps $J_n^{\mathcal F\mathcal F}$, $J_n^{\mathcal F\mathcal P}$, $J_n^{\mathcal P\mathcal F}$, and $J_n^{\mathcal P\mathcal P}$, each one connecting the spaces indicated in the superscript---for instance, $J_n^{\mathcal F\mathcal P}$ connects $\mathcal P_n$ with $\mathcal F_n$. As $\mathcal F_n$ and $\mathcal P_n$ transform with the same irreducible representation of SO(4), each of these four components transform with the same representation. To be invariant, they all must commute with all the elements of the SO(4) group. That being the case, Schur's lemma~\cite{Kirillov76} implies that the four maps are proportional to the identity one, and hence they must respect the labels $l$ and $m$ as well~\cite{CMV10}. Therefore,
one can conclude that the matrix representation of $J_n$ must be block diagonal, with each
$2\times 2$ block~$j_n$ acting just on the linear span of $\{\bar f_{\boldsymbol n}, \bar\pi_{\bar f_{\boldsymbol n}}\}$,
with fixed $\boldsymbol n$, although its explicit form depends only on $n$ (or, equivalently, on the eigenvalue of the Laplace-Beltrami operator), but not on $l$ and $m$.

If we consider now a generic invariant complex structure~$J$, we can assign annihilation and creation variables to it. This new basis on phase space can be related to the old one, formed by the corresponding variables of the complex structure $J_0$, by a symplectomorphism~$\mathcal K$ with a block diagonal matrix representation of the same type as $J$~\cite{CCMV06,CMV10}. We parametrize its blocks~$\mathcal K_n$ in the form
\begin{equation}\label{eq:K_n}
\mathcal K_n =
\begin{pmatrix}
\kappa_n & \lambda_n \\
\lambda_n^* & \kappa_n^*
\end{pmatrix},
\end{equation}
where we have used again that all the maps under consideration are real, and we have $|\kappa_n|^2-|\lambda_n|^2=1$, so that $\mathcal K$ preserves the symplectic structure. Then, employing a change of basis of this kind, the relation between a generic invariant complex structure and the one of reference, $J_0$, is $J=\mathcal KJ_0\mathcal K^{-1}$ or, block by block, $ j_n=\mathcal K_n j_0\mathcal K_n^{-1}$.

Let us then assume that we are given an invariant complex structure~$J$, distinct from $J_0$, which allows for a natural implementation of the dynamics. We will now show that any such complex structure gives rise to a representation which is unitarily equivalent to the one defined by $J_0$. The necessary and sufficient condition for the unitary implementation of the symplectomorphism~$\mathcal U$ with respect to this new complex structure is that the transformation~$\mathcal K^{-1}\mathcal U \mathcal K$ (obtained with a change of basis as discussed above) be unitarily implementable with respect to the original complex structure~$J_0$~\cite{CCMV06}. Of course, the matrix representation of $\mathcal K^{-1}\mathcal U\mathcal K$ also consists of blocks, which we can write as
\begin{equation}
\mathcal K_n^{-1}\mathcal U_n\mathcal K_n =
\begin{pmatrix}
\alpha^J_n & \beta^J_n \\
\big(\beta^J_n\big)^* & \big(\alpha^J_n\big)^*
\end{pmatrix},
\end{equation}
where
\begin{equation}
\beta^J_n = \big(\kappa_n^*\big)^2\beta_n-\lambda_n^2\beta_n^*+2i\kappa_n^*\lambda_n\Im(\alpha_n).
\end{equation}
The symbol $\Im$ stands for the imaginary part.
We will not need an explicit expression for $\alpha^J_n$, because the unitary implementation of $\mathcal U$ with respect to $J$ relies only on the square summability of the sequence $\{n\beta^J_n\}_n$. In the following, we will show that this square summability (at all times $\eta$) implies indeed that of $\{n\lambda_n\}_n$, which is precisely the necessary and sufficient condition for the equivalence of the invariant complex structure $J$ and of the reference one, $J_0$, the relation between them being given by a unitary transformation.

If $n\beta^J_n(\eta,\eta_0)$ forms a square-summable sequence, at least for $\eta$ in a certain time interval~$I$ around the arbitrary initial time $\eta_0$, the same will happen with $n\beta^J_n(\eta,\eta_0)/(\kappa_n^*)^2$, because $|\kappa_n|^2=1+|\lambda_n|^2\geq 1$. We can subtract to each term of this latter sequence the contributions that we know that are square summable (i.e., any term of the order of $n^{-1}$ or smaller), so as to obtain another square-summable sequence. From Eq.~\eqref{eq:alpha_n}, it is not difficult to see that $\alpha_n=e^{\omega_n\Theta_n}+O(n^{-2})$, and introducing the asymptotic limit of $W_n$ \eqref{eq:asymptoticW_n} in Eq.~\eqref{eq:Theta_n}, one concludes that the sequence
\begin{equation}
\bigg\{n\frac{\lambda_n}{\kappa_n^*}\sin\!\bigg[\omega_n(\eta-\eta_0)+\int_{\eta_0}^\eta\frac{s(\bar\eta)}{2\omega_n}d\bar\eta\bigg]\bigg\}_n
\end{equation}
must be square summable in $I$. Let $Z(\eta)$ be the sum of its terms squared. If $Z$ were integrable in $I$, we might write
\begin{align}
\Lambda_{n_0}\sum_{n=n_0}^Mn^2\left|\frac{\lambda_n}{\kappa_n}\right|^2 \leq \sum_{n=n_0}^Mn^2\left|\frac{\lambda_n}{\kappa_n}\right|^2\int_Id\eta\,\sin^2\!\left[\omega_n(\eta-\eta_0)+\int_{\eta_0}^\eta\frac{s(\bar\eta)}{2\omega_n}
d\bar\eta\right] \leq \int_Id\eta\,Z(\eta), \label{eq:inequality}
\end{align}
where $M$ is any integer bigger than $n_0$, and this latter integer is chosen so that the integral over $I$ in the second term of the above inequality is bounded from below by a strictly positive quantity $\Lambda_{n_0}$ for all $n>n_0$~. This is always possible because, when $n\rightarrow\infty$, the last summand in the argument of the sine tends to zero and the integral can be seen to approach half the Lebesgue measure of the interval $I$~\cite{CMV10}. Hence, the sequence $\{n\lambda_n/\kappa_n\}_n$ would turn out to be square summable, as the partial sums of its square elements form a non-decreasing sequence that would be bounded from above.

Even if $Z$ were not integrable, since it is the limit of a sequence of (partial sums of) measurable functions, and therefore it is measurable itself, Luzin's theorem~\cite{KF99} ensures that, for all $\delta>0$, there exists a set~$I_\delta\subset I$ such that the measure of $I\setminus I_\delta$ is less than $\delta$ and the restriction $Z_\delta$ of $Z$ to $I_\delta$ is a continuous function. Using this fact, we can restrict the integrals of expression $\eqref{eq:inequality}$ to the set~$I_\delta$. A lower bound for the integral of the squared sine is then $\Lambda_{n_0}-\delta$, which is still strictly positive provided that we choose a sufficiently small $\delta$. Therefore, the sequence formed by $n\lambda_n/\kappa_n$ turns out to be square summable in this situation as well, for similar reasons to those explained above.

Bearing in mind that $|\kappa_n|^2-|\lambda_n|^2=1$, it is easy to see that the square summability of $\{n\lambda_n/\kappa_n\}_n$ implies that $\lambda_n$ tends to zero and $\kappa_n$ to the unity when $n$ approaches infinity. Consequently, the sequence formed by $\kappa_n$ is bounded from above, and thus $1/\kappa_n$ is bounded from below. It then follows straightforwardly that the sequence $\{n\lambda_n\}_n$ must be square summable. In this way, we conclude that the symplectomorphism~$\mathcal K$, relating two SO(4)-invariant complex structures, is indeed unitarily implementable. Therefore, all invariant complex structures which allow a unitary dynamics are unitarily equivalent. This proves that the criteria of symmetry invariance and unitary evolution that we have imposed select a unique equivalence class of Fock representations. Again, we see that the modifications of the field dynamics that appear in the equations of motion \eqref{eq:momentum} and \eqref{eq:theequation} with respect to those analyzed in Ref.~\cite{CMV10} do not alter the result of uniqueness of the Fock representation, essentially because the asymptotic behavior of the Bogoliubov coefficients is changed only at the order of $n^{-2}$ or smaller.

\subsection{Uniqueness of the field description}\label{subsec:uniqueness}

We have obtained a Fock quantization for the perturbation of the scalar field in terms of the modes $\bar f_{\boldsymbol n}$ and their conjugate momenta. This quantization is characterized by a complex structure with SO(4) symmetry and unitary dynamics. In addition, we have proven that any other complex structure with the same properties leads to a unitarily equivalent representation of the CCR's associated with our choice of a canonical pair of field variables. However, there is still the possibility that a different choice of field variables might result in a different quantization. Specifically, our field description was reached after performing a particular scaling by a time-dependent function. The question arises of whether, by performing a different time-dependent scaling, one might attain a distinct field description in which there might exist an invariant Fock representation of the CCR's leading to a unitary evolution and such that the corresponding quantum theory could not be related with the one that we have constructed by a unitary transformation. It is worth recalling that a time-dependent scaling of the field can be completed into a canonical transformation by introducing the inverse scaling of the field momentum. One may also allow for a time-dependent, linear contribution of the field configuration to the new momentum while respecting the linearity and locality of the system. Since this canonical transformation depends on time, the dynamics in the new field description differs from the original one. Therefore, in principle, the criterion of unitary implementability of the evolution imposes different conditions on the diverse field descriptions that can be obtained in this way. In this section, we want to prove that, in fact, there is no freedom to change the field description if the criteria of SO(4) invariance and unitary dynamics are to be fulfilled.

Mode by mode, the class of time-dependent canonical transformation that we are considering has the form:
\begin{subequations}\label{eqs:breve}
\begin{align}
\breve f_{\boldsymbol n}(\eta) &= F(\eta)\bar f_{\boldsymbol n}(\eta), \\
\breve\pi_{\breve f_{\boldsymbol n}}(\eta) &= \frac1{F(\eta)}\bar\pi_{\bar f_{\boldsymbol n}}(\eta)+G(\eta) \bar f_{\boldsymbol n}(\eta).
\end{align}
\end{subequations}
It is important to emphasize that the time-dependent factors in this linear transformation are the same for all the modes. Thus, the canonical transformation is local in terms of the configuration and momentum field variables. Besides, we assume that the functions $F$ and $G$ are real, twice differentiable, and that $F$ does not vanish anywhere, so that we neither spoil the differential structure formulation of the field theory nor introduce artificial singularities. If $F$ and $G$ were constant, the quantum representation for the new field variables would be the same as for the barred ones, with the new canonical pair straightforwardly obtained from the original one by mere constant linear combinations. Incidentally, this allows us to fix the initial conditions so that the two canonical pairs coincide at the time $\eta_0$, that is, we can set $F(\eta_0)=1$ and $G(\eta_0)=0$ with no loss of generality.

Thus, we are interested only in time-dependent transformations, which affect the dynamics. In fact, if one defines annihilation and creation variables combining the new variables in a way analogous to Eq.~\eqref{eq:creation}, their evolution can be still expressed as in Eq.~\eqref{eq:U_n}, except that now the Bogoliubov coefficients are \cite{CMV07,CMOV11}:
\begin{subequations}\label{eqs:breveU}
\begin{align}
\breve\alpha_n(\eta,\eta_0) &= F_+(\eta)\alpha_n(\eta,\eta_0)+F_-(\eta)\beta_n^*(\eta,\eta_0)
+\frac{i}{2\omega_n}G(\eta)[\alpha_n(\eta,\eta_0)+\beta_n^*(\eta,\eta_0)], \\
\breve\beta_n(\eta,\eta_0) &= F_+(\eta)\beta_n(\eta,\eta_0)+F_-(\eta)\alpha_n^*(\eta,\eta_0)
+\frac{i}{2\omega_n}G(\eta)[\beta_n(\eta,\eta_0)+\alpha_n^*(\eta,\eta_0)],
\end{align}
\end{subequations}
where $2F_\pm=F\pm1/F$. Remarkably, the corresponding symplectomorphism $\breve{\mathcal U}$ is not unitarily implementable with respect to $J_0$ in any time interval. Furthermore, it has been shown~\cite{CMOV11} that, actually, there exists no SO(4)-invariant complex structure in whose corresponding representation $\breve{\mathcal U}$ is implemented by a unitary operator, unless the transformation~\eqref{eqs:breve} is the trivial identity. We briefly sketch the proof here.

Let us suppose that there is such an invariant complex structure~$J$. It will be related to $J_0$ by a symplectomorphism~$\mathcal K$, as we have seen above. Then, if the dynamics of the new field description of the system admits a unitary implementation with respect to $J$, $\mathcal K^{-1}\breve{\mathcal U}\mathcal K$  must be unitarily implementable with respect to $J_0$. Parametrizing the blocks~$\mathcal K_n$ of the matrix representation of $\mathcal K$ as in Eq.~\eqref{eq:K_n}, direct arguments show the square summability of the sequence of elements $n\breve\beta^J_n(\eta,\eta_0)$ (at all times $\eta$ in the considered interval $I$), where the Bogoliubov coefficients~$\breve\beta^J_n$ are given by
\begin{equation}
\breve\beta^J_n = \left(\kappa_n^*\right)^2\breve\beta_n-\lambda_n^2\breve\beta_n^*+2i\kappa_n^*\lambda_n\Im(\breve\alpha_n).
\end{equation}
If this is the case, the coefficients $\breve\beta^J_n/(\kappa_n^*)^2$ must clearly tend to zero as $n\rightarrow\infty$. Using Eqs.~\eqref{eqs:breveU}, together with the asymptotic limits of $\alpha_n$ and $\beta_n$, and the fact that $\omega_n=n+O(n^{-1})$, we arrive at the conclusion that the sequence of elements
\begin{align}\label{eq:lim}
\bigg[e^{in(\eta-\eta_0)}-\bigg(\frac{\lambda_n}{\kappa_n^*}\bigg)^2e^{-in(\eta-\eta_0)}\bigg]F_-(\eta) 
-2i\frac{\lambda_n}{\kappa_n^*}\sin\big[n(\eta-\eta_0)\big]F_+(\eta)
\end{align}
must tend to zero at large $n$. We can fix a time of the form $\eta=\eta_0+2\pi q/p$ in the interval $I$, with $q$ in $\mathbb N$ and $p$ in $\mathbb N^+$, so as to simplify the above expression, and consider only the subsequence formed by the terms with $n=mp$ ($m\in\mathbb N^+$). With this in mind, the real and imaginary parts of the quantities~\eqref{eq:lim} read
\begin{subequations}
\begin{align}
& \bigg[1-\Re\bigg(\frac{\lambda_{mp}^2}{(\kappa_{mp}^*)^2}\bigg)\bigg]F_-\bigg(\eta_0+\frac{2\pi q}p\bigg), \\
& -\Im\bigg(\frac{\lambda_{mp}^2}{(\kappa_{mp}^*)^2}\bigg)F_-\bigg(\eta_0+\frac{2\pi q}p\bigg).
\end{align}
\end{subequations}
Then, both sequences of elements must approach zero as $m\rightarrow\infty$. And this must be so for all possible choices of the positive integers $q$ and $p$. But this is only possible if $F_-(\eta_0+2\pi q/p)=0$, since it can be proven that the other factors cannot tend to zero simultaneously~\cite{CMV07,CMOV11}. Given that the set of points $\eta_0+2\pi q/p$ under consideration is dense in the time interval $I$, and recalling that the function $F$ is continuous [with $F(\eta_0)=1$], we are led to conclude that $F(\eta)$ has to be the unit function.

Substituting $F(\eta)=1$ in the terms~\eqref{eq:lim}, it becomes clear that $\lambda_n/\kappa_n^*$ tends to zero as $n$ increases (the sine function cannot tend to zero at all times in $I$ \cite{CMOV11}). One can then deduce from the asymptotic behavior of the sequence $\{n\beta^J_n/(\kappa_n^*)^2\}_n$ that the terms
\begin{equation}
\label{z1}
G(\eta)-4n\frac{\lambda_n}{\kappa_n^*}\sin\big[n(\eta-\eta_0)\big]e^{-in(\eta-\eta_0)}
\end{equation}
have also a zero limit. Considering the imaginary part of this sequence, one can show that  $n \lambda_n/\kappa_n^*$ must also tend to zero (again because a sequence of functions of the type $\sin\big[n(\eta-\eta_0)\big] \sin\big[n(\eta-\eta_0)+\delta_n\big]$ cannot tend to zero at all times \cite{CMOV11}). It follows that the sequence of terms (\ref{z1}) can have a zero limit if and only if the function $G$ vanishes identically. Thus, the transformation~\eqref{eqs:breve} is actually the identity; only in this case can $\breve{\mathcal U}$ be implemented in a unitary way.

In summary, a time-dependent scaling of the field perturbation other than the Mukhanov scaling prevents the unitary implementation of the evolution. Note that, for instance, this is not the scaling adopted in Ref. \cite{HH85}. The requirements of unitary quantum dynamics and SO(4) symmetry select a preferred choice of canonical pair of field variables and, as we have proven, a unique unitary equivalence class of Fock representations for them.

\subsection{An equivalent quantization}\label{subsec:another}

While implementing the canonical transformation \eqref{eqs:transformation} in Sec.~\ref{sec:classical}, we took care that the coupling terms between the scalar field perturbation and its canonical momentum in the new Hamiltonian decreased sufficiently fast in the limit of large $n$. This issue was key to the unitary implementation of the dynamics in the Fock representation introduced in Sec.~\ref{subsec:unitary}. Of course, it is possible to perform a canonical transformation that removes those coupling terms completely. The price to pay is that each mode of the perturbation has to be treated separately, instead of transforming the perturbation as a whole. In other words, one has to perform a canonical transformation that is defined in a different way for each mode, and therefore cannot be defined locally in terms of the original field and its momentum. We address now this alternate kind of transformations. It is important, however, not to spoil the linearity of the system, at least. For this reason, we consider linear canonical transformations of the type \eqref{eqs:breve}, but with mode-dependent functions $F_n$ and $G_n$. It is easy to see that the election
\begin{subequations}
\begin{equation}
F_n = \frac1{\sqrt{\bar E^n_{\pi\pi}}}
\end{equation}
drops the term linear in the first derivative from the equation of motion, which can be written in the form~\eqref{eq:theequation}, with $r_n=0$, whereas $s_n$ is still given by Eq.~\eqref{eq:s_n}, up to corrections of the order of $n^{-2}$. In other words, the new configuration field variables obey equations very similar to those for the modes of a Klein-Gordon field, except that the mass is a time function with mode-dependent, subdominant contributions. We complete the transformation with
\begin{equation}
G_n = \frac1{\sqrt{\bar E^n_{\pi\pi}}}\left(\bar E^n_{f\pi}-\frac12\frac{\dot{\bar E}^n_{\pi\pi}}{\bar E^n_{\pi\pi}}\right).
\end{equation}
\end{subequations}
Then, the new momentum variables satisfy exactly the relation $\breve\pi_{\breve f_{\boldsymbol n}}=\dot{\breve f}_{\boldsymbol n}$. In this way, we totally eliminate the configuration-momentum coupling terms of the field perturbations in the Hamiltonian.

With the system in this form, the results for the case of a scalar field with time-dependent mass on $S^3$~\cite{CMV10} can be applied in a straightforward way. Accordingly, there exists a unique unitary equivalence class of Fock representations of the CCR's for the new canonical pair of field variables which possesses the desired properties of unitary dynamics and SO(4) symmetry. A representative of that class is the quantization that can be constructed in terms of the annihilation and creation variables defined by combining the new configuration and momenta variables in the same way as we did with the barred ones in Eq.~\eqref{eq:creation}. A block-diagonal symplectomorphism~$\mathcal K$ relates the new annihilation and creation variables with the ones previously defined. Of course, its blocks~$\mathcal K_n$ can again be parametrized as in Eq.~\eqref{eq:K_n}, and a calculation shows that
\begin{equation}
\lambda_n = \frac12\left(F_n-\frac1{F_n}\right)+\frac i{2\omega_n}G_n.
\end{equation}
Recalling that $\bar E^n_{\pi\pi}=1+O(n^{-2})$ and that $\bar E^n_{f\pi}$ decreases asymptotically as $n^{-2}$, according to Eqs.~\eqref{eqs:E^n}, it is easy to check that the sequence $\{n\lambda_n\}_n$ is square summable. This is precisely the condition for $\mathcal K$ to be unitarily implementable with respect to $J_0$. Therefore, the new quantization is unitarily equivalent to the one discussed in the preceding subsections.

Thus, we obtain an equivalent quantum theory for the scalar field perturbation even though we transform each mode individually, a fact which prevents us from employing directly the uniqueness results explained above about changes of field description resulting from canonical transformations that vary on time. It is crucial to emphasize that the uniqueness of the quantization is granted only because we have started from the massive scalar field and we have considered time-dependent scalings of it, showing that there exists exclusively one which satisfies our criteria. For that particular field description, we have demonstrated the uniqueness of the Fock representation, and then we have shown that one can carry out a unitary transformation in that quantum theory so as to absorb all coupling terms between configuration and momentum field variables in the system. Had we started with the quantum representation introduced in this subsection and admitted the possibility of performing time-dependent canonical transformations mode by mode, in principle, we would not have been able to remove the ambiguity in the choice of a specific field description.

\section{Gauge-invariant variables}\label{sec:Bardeen}

In the study of cosmological perturbations, gauge-invariant quantities are usually employed to describe the physics in a consistent manner, independent of the identification of the spacetime and its matter content modulo diffeomorphisms, and insensitive to the specification of a particular gauge. In this section, we will construct a canonical pair of such quantities in terms of the canonical pair of field variables used so far, and discuss their relation with the set of well-known gauge-invariant variables introduced by Bardeen~\cite{Bardeen80}.

Let us consider a gauge transformation of the type $x^\mu \mapsto x^\mu+\epsilon\xi^\mu$, where $\epsilon\xi^\mu$ represents a small displacement that we will treat as a perturbation ($\epsilon$~being the perturbative parameter introduced in Sec.~\ref{sec:classical}). The covariant counterpart of $\xi^\mu$ can be decomposed using hyperspherical harmonics:
\begin{align}
\xi_0 = \sqrt2\,\pi\sigma^2 N_0\sum_{\boldsymbol n}T_{\boldsymbol n}\tilde{Q}^{\boldsymbol n}, \qquad
\xi_i = \sqrt2\,\pi\sigma^2e^{\alpha}\sum_{\boldsymbol n}L_{\boldsymbol n}(\tilde{P}_i)^{\boldsymbol n},
\end{align}
with $T_{\boldsymbol n}$ and $L_{\boldsymbol n}$ being functions only of time. Genuine vector harmonics do not need to be included, as they would only affect the genuine vector perturbations of the metric~\cite{Bardeen80}, which we have ignored.

Under the considered transformation, the background spacetime is understood to remain the same, but the perturbative coefficients appearing in Eqs.~\eqref{eqs:variables} change. Including corrections up to the order of $\epsilon$ at most, the change is:
\begin{subequations}
\begin{align}
a_{\boldsymbol n} &\mapsto a_{\boldsymbol n}+e^{-\alpha}\left(\dot\alpha T_{\boldsymbol n}+\tfrac13L_{\boldsymbol n}\right), \\
b_{\boldsymbol n} &\mapsto b_{\boldsymbol n}-\tfrac13e^{-\alpha}L_{\boldsymbol n}, \\
f_{\boldsymbol n} &\mapsto f_{\boldsymbol n}+e^{-\alpha}\dot\varphi\,T_{\boldsymbol n}, \\
g_{\boldsymbol n} &\mapsto g_{\boldsymbol n}+e^{-\alpha}\dot T_{\boldsymbol n}, \\
k_{\boldsymbol n} &\mapsto k_{\boldsymbol n}-N_0e^{-\alpha}\big(\omega_n^2T_{\boldsymbol n}+\dot L_{\boldsymbol n}-\dot\alpha L_{\boldsymbol n}\big).
\end{align}
\end{subequations}
Recall that the dot stands for the derivative with respect to the conformal time. It is easy to combine the above variables to construct gauge-invariant ones at the considered perturbative order. Two independent examples of these are
\begin{subequations}\label{eqs:invariant}
\begin{align}
\mathcal E^m_{\boldsymbol n} &= \frac{e^{-2\alpha}}{E_0}\left[-\dot\varphi^2g_{\boldsymbol n}+\dot\varphi\dot f_{\boldsymbol n}+(3\dot\alpha\dot\varphi+e^{2\alpha}m^2\varphi)f_{\boldsymbol n}\right] \\
v^s_{\boldsymbol n} &= \frac{1}{\omega_n}\left[\frac{\omega_n^2}{\dot\varphi}f_{\boldsymbol n}+\left(\frac{k_{\boldsymbol n}}{N_0}-3\dot b_{\boldsymbol n}\right)\right],
\end{align}
\end{subequations}
where we have introduced a function proportional to the energy density of the background scalar field, \begin{equation}E_0=\frac12(e^{-2\alpha}\dot\varphi^2+m^2\varphi^2).\end{equation}
The quantities that we have defined correspond to two of the gauge-invariant scalar quantities introduced by Bardeen, namely, the energy density and matter velocity perturbations~\cite{Bardeen80}. In the longitudinal gauge, and using the variables~\eqref{eqs:tilde} and \eqref{eqs:transformation}, their expressions reduce to
\begin{subequations}
\begin{align}
\mathcal E^m_{\boldsymbol n} &= \frac{e^{-5\bar\alpha}}{E_0}\left[\pi_\varphi\bar\pi_{\bar f_{\boldsymbol n}}-e^{-2\bar\alpha}\left(2\bar\pi_{\bar\alpha}\pi_\varphi-e^{6\bar\alpha}m^2\tilde\varphi\right)\bar f_{\boldsymbol n}\right]+O(\epsilon^2), \\
v^s_{\boldsymbol n} &= \sqrt{n^2-1}\frac{e^{\bar\alpha}}{\pi_\varphi}\bar f_{\boldsymbol n}+O(\epsilon^2).
\end{align}
\end{subequations}
Obviously, any combination of these invariant quantities and the background variables is also gauge-invariant. Let us define the functions
\begin{subequations}\label{eqs:Psi_n1}
\begin{align}
\Psi_{\boldsymbol n} &= \frac{1}{\sqrt{n^2-4}}\frac{e^{5\alpha}}{\pi_\varphi}E_0\mathcal E^m_n, \\
\Pi_{\Psi_{\boldsymbol n}} &= -\sqrt{\frac{n^2-4}{n^2-1}}e^{-\alpha}\pi_\varphi v^s_{\boldsymbol n}
-\frac{e^{-2\alpha}}{\pi_\varphi}(2\pi_\alpha\pi_\varphi-e^{6\alpha}m^2\varphi)\Psi_{\boldsymbol n}.
\end{align}
\end{subequations}
These quantities satisfy the relation $\dot\Psi_{\boldsymbol n}=\Pi_{\Psi_{\boldsymbol n}}+O(\epsilon)$, as can be seen from the equation
\begin{equation}\label{eq:E-v}
\frac{d}{d\eta}\!\left(e^{3\alpha}E_0\mathcal E^m_{\boldsymbol n}\right) \approx -\frac{n^2-4}{\sqrt{n^2-1}}e^{-3\alpha}\pi_\varphi^2v^s_{\boldsymbol n}+O(\epsilon),
\end{equation}
which holds for any isotropic matter content~\cite{Bardeen80}, as it happens to be our case. In terms of $\bar f_{\boldsymbol n}$ and $\bar\pi_{\bar f_{\boldsymbol n}}$, they are given by
\begin{subequations}\label{eqs:Psi_n2}
\begin{align}
\Psi_{\boldsymbol n} &= \frac{1}{\sqrt{n^2-4}}\left(\bar\pi_{\bar f_{\boldsymbol n}}+\chi\bar f_{\boldsymbol n}\right)+O(\epsilon^2), \\
\Pi_{\Psi_{\boldsymbol n}} &= \frac\chi{\sqrt{n^2-4}}(\bar\pi_{\bar f_{\boldsymbol n}}+\chi\bar f_{\boldsymbol n})-\sqrt{n^2-4}\bar f_{\boldsymbol n}+O(\epsilon^2),
\end{align}
\end{subequations}
where
\begin{equation}
\chi = -\frac{e^{-2\bar\alpha}}{\pi_\varphi}(2\bar\pi_{\bar\alpha}\pi_\varphi-e^{6\bar\alpha}m^2\tilde\varphi).
\end{equation}
Using these expressions, it is easy to see that $\Psi_{\boldsymbol n}$ and $\Pi_{\Psi_{\boldsymbol n}}$ are in fact canonically conjugate on the complete phase space of our reduced system (including the homogeneous degrees of freedom), as the notation suggests. In addition, the functions $\Psi_{\boldsymbol n}$ satisfy the same kind of dynamical equations as the modes of a scalar field with time-dependent mass:
\begin{equation}
\ddot\Psi_{\boldsymbol n}
+\left(n^2-4-3e^{-4\bar\alpha}\pi_\varphi^2-\dot\chi-\chi^2\right)\Psi_{\boldsymbol n} = O(\epsilon) .
\end{equation}
Therefore, we can construct a Fock representation with SO(4) symmetry and unitary dynamics by defining annihilation and creation variables in the usual way \cite{CMV10}:
\begin{equation}\label{eq:creation2}
\begin{pmatrix}
a_{\Psi_{\boldsymbol n}} \\
a_{\Psi_{\boldsymbol n}}^*
\end{pmatrix}
= \frac1{\sqrt{2\omega_n}}
\begin{pmatrix}
\omega_n & i \\
\omega_n & -i
\end{pmatrix}
\begin{pmatrix}
\Psi_{\boldsymbol n} \\
\Pi_{\Psi_{\boldsymbol n}}
\end{pmatrix}.
\end{equation}
Any other representation with the same properties must be unitarily equivalent to this one \cite{CMV10}.

The canonical transformation that leads to the variables $(\Psi_{\boldsymbol n},\Pi_{\Psi_{\boldsymbol n}})$ is more general than the ones we have considered in Sec.~\ref{subsec:another}, for the new coordinates have also contributions from the old momenta. Nevertheless, it does not mix the labels $n$, $l$, and $m$, so the symplectomorphism~$\mathcal K$ connecting the two bases of annihilation and creation variables is still block diagonal, with blocks~$\mathcal K_n$ of the form~\eqref{eq:K_n}. The antilinear part of this symplectomorphism is characterized by the coefficients
\begin{equation}\label{eq:lambda_n}
\lambda_n = \frac{i}{2}\frac{\chi^2+3}{\sqrt{n^2-1}\sqrt{n^2-4}}.
\end{equation}
Therefore, the transformation can be implemented by a unitary operator in the quantum theory, as the sequence of elements $n\lambda_n$ (where we are taking into account the degeneracy of the eigenspaces of the Laplace-Beltrami operator) is certainly square summable at all times, \emph{provided} $\pi_\varphi$ does not vanish and hence $\chi$ is well defined. Thus, with this caveat, we could alternatively quantize the perturbation in terms of the gauge-invariant variables $(\Psi_{\boldsymbol n},\Pi_{\Psi_{\boldsymbol n}})$ and obtain a quantum theory which is unitarily equivalent to the one which is picked up by our criteria of symmetry invariance and unitary evolution, circumventing in this way any gauge dependence.

\section{Conclusion}

With our analysis, we have shown in a specific example that the uniqueness results for the quantization of a scalar field with time-dependent mass in a compact spatial manifold~\cite{CMV10,CMOV11} can be generalized to situations in which the dynamical equations of the system have additional
corrective terms, as long as they present a suitable subdominant behavior in the large-$n$ limit or, equivalently, in the asymptotic limit of large eigenvalues $\omega_n$ of the Laplace-Beltrami operator. In particular, we have studied the quantization of the scalar perturbations of a massive field coupled to the gravitational field in an FLRW model. We have considered the case in which the spatial sections are homeomorphic to three-spheres, as a standard example of compact topology, in which infrared divergences do not appear. Nonetheless, the conclusions should be generalizable to other compact topologies, like e.g. the case of flat sections with the topology of a three-torus, in view of the results developed recently in Ref.~\cite{lastCMOV}.

The local gauge degrees of freedom have been removed with two alternate sets of gauge-fixing conditions: in the main body of the article, we have adopted the longitudinal gauge, in which the shift vector vanishes and the three-metric of the spatial sections is conformal to that of the three-sphere, whereas in the Appendix we have imposed the homogeneity of the spatial sections. After these gauge fixations, the remaining local degrees of freedom can be identified with those of the perturbation of the matter field. The uniqueness results of Refs.~\cite{CMV10,CMOV11} can be extended to both of these gauge-fixed systems. The modifications in the second-order field equation of the Klein-Gordon type, in the form of a dissipative term and of a mode-dependent correction to the mass, as well as the changes in the relation between the field momentum and the time derivative of the field configuration, turn out to be of the order of $n^{-2}$ or smaller. These modifications do not alter the dynamical behavior of the modes in the ultraviolet limit (and, {\it a fortiori}, of the Bogoliubov coefficients) up to corrections $O(n^{-2})$, which is precisely the order that can be neglected in the proof of the uniqueness of the Fock quantization of the system, both concerning the adoption of a field description and of a Fock representation for it.

In such a quantization, the homogeneous degrees of freedom have been treated classically, thus neglecting the  quantum effects arising from them. This is usually expected to be a good approximation away from the Planck regime. Since the system has been linearized, the perturbation of the scalar field admits a standard Fock quantization. We have restricted to complex structures that share the SO(4) symmetry of the spatial sections---hence defining an SO(4)-invariant vacuum state---and permit a unitary implementation of the field dynamics. These two properties suffice to select a unique unitary equivalence class of representations for a suitable choice of the canonical pair of field variables. A representative of this class is the quantization constructed with the complex structure that would be natural in the case of a massless field on the three-sphere.

Furthermore, if one explores other field descriptions with distinct canonical pairs, attainable by means of a time-dependent scaling of the field after completing it into a linear canonical transformation, no additional Fock representation can be found that meets the two requirements simultaneously. The correct scaling of the field has been implemented in Sec.~\ref{subsec:gauge} as a canonical transformation on the complete phase space of the system, \emph{including} the homogeneous degrees of freedom. In addition, we have determined more general, mode-by-mode time-dependent linear canonical transformations that allow us to reformulate the system in the genuine form of what would be a scalar field with time-varying mass on the three-sphere, with no subdominant corrections, and for which one can apply previous uniqueness results about the Fock representation of the CCR's (based also on the same criteria)~\cite{CMV10}. In this way, we have obtained two new quantum Fock descriptions of the perturbations. Actually, one of these formulations is made in terms of gauge-invariant quantities: the energy density and matter velocity perturbations introduced by Bardeen~\cite{Bardeen80}, whose definition is independent of the gauge. In particular, our analysis has led us to establish a canonical structure for gauge-invariant variables. We have proven that the two mentioned, alternate Fock quantizations of the system are unitarily equivalent to the one picked up by the criteria of symmetry invariance and unitary evolution for the scalar perturbations of the massive field. We also notice that the canonical pair of gauge invariants that we have found differs from others studied in previous works~\cite{Langlois94,CB11}. Besides, let us mention that it is possible to show that this pair can be related to the Mukhanov-Sasaki variable (for this non-flat spatial case) together with a suitable conjugate momentum by means of a transformation that is unitarily implementable, therefore leading to a quantization which is equivalent as well \cite{lang}.

It is worth emphasizing that it is only for the massive scalar field that we can naturally consider time-dependent scalings that absorb part of the time variation of the background, while respecting the locality and the linearity of the system. Mode-by-mode transformations are defined in general in a non-local way. Thus, starting with the massive field we can select a unique scaling and choose a privileged canonical pair by demanding our criteria of vacuum invariance and unitary dynamics. For the corresponding CCR's, the same criteria provide a unique, preferred Fock representation. And, from this unique Fock quantization, a unitary transformation leads to an alternate quantization, adapted to the mentioned canonical pair of Bardeen's gauge invariants. The consistency in the application of our criteria to select the quantum theory is complete, for we have further shown that the Fock quantization determined in this way is precisely the unique one which implements the evolution of those gauge invariants in a unitary way, while preserving the SO(4) symmetry. Note also that it is the fact that we have begun with the quantization of the massive scalar field what allows us to choose exactly that canonical pair of Bardeen's potentials and not any other one.

The robustness that these uniqueness results confer to the physical predictions of the quantization is remarkable. The unitary implementation of the dynamics, together with the symmetries of the spatial sections, prove to be powerful criteria to select not only preferred canonical field variables, but also a preferred Fock representation for them. If one accepts these criteria, the conclusions of the present work have implications in the study of structure formation and cosmological inflation, since they pick up a quantization of the (scalar) cosmological perturbations. One particularly interesting application is found in the so-called hybrid quantization approach to cosmology. This approach, which combines the Fock quantization of the inhomogeneities with the polymer quantization of the zero modes of the system~\cite{MGM08}, allows one to include local degrees of freedom in Loop Quantum Cosmology~\cite{LQC,lqc,gamm}. The completion of this program in the case of the scalar perturbations of an FLRW spacetime will be the subject of future research~\cite{FMO12}.

\begin{acknowledgments}
This work was supported by the research grants Nos.
MICINN/MINECO FIS2011-30145-C03-02, MICINN FIS2008-06078-C03-03, and CPAN
CSD2007-00042 from Spain, and CERN/FP/116373/2010 from Portugal. J.O. acknowledges
CSIC for financial support under the grant No.\ JAE-Pre\_08\_00791, and M.F.-M. acknowledges CSIC and the European Social Fund for support under the grant No.\ JAEPre\_2010\_01544.
\end{acknowledgments}

\appendix

\section{A different gauge}

In this appendix, we choose gauge-fixing conditions different from those given by Eqs.~\eqref{eqs:gauge1}. More precisely, we impose
\begin{equation}\label{eq:gauge2}
a_{\boldsymbol n}=0=b_{\boldsymbol n},
\end{equation}
so that the perturbation of the three-metric vanishes. In order to check the validity of these conditions, we calculate their Poisson brackets with the constraints that are going to be fixed,
\begin{equation}
\epsilon^4\det
\begin{pmatrix}
\{a_{\boldsymbol n}, H_{\_1}^{\boldsymbol n}\} & \{b_{\boldsymbol n}, H_{\_1}^{\boldsymbol n}\} \\
\{a_{\boldsymbol n}, H_{|1}^{\boldsymbol n}\} & \{b_{\boldsymbol n}, H_{|1}^{\boldsymbol n}\}
\end{pmatrix}
= \frac13e^{-4\alpha}\pi_\alpha,
\end{equation}
which becomes zero exclusively when $\pi_\alpha=0$. As far as that point is eluded (or, rather, that section on phase space), the fixation of the gauge is admissible. We still have to require its stability under the dynamics, which implies
\begin{subequations}
\begin{align}
0 = \{a_{\boldsymbol n},H\} &\approx -N_0e^{-3\alpha}\big(\pi_{a_{\boldsymbol n}}+g_{\boldsymbol n}\pi_\alpha\big) -\frac13e^{-\alpha}k_{\boldsymbol n}+O(\epsilon), \\
0 = \{ b_{\boldsymbol n},H\} &\approx N_0e^{-3\alpha}\frac{n^2-1}{n^2-4}\pi_{b_{\boldsymbol n}}  +\frac13e^{-\alpha}k_{\boldsymbol n}+O(\epsilon).
\end{align}
\end{subequations}
The latter equations can be solved for the Lagrange multipliers $g_{\boldsymbol n}$ and $k_{\boldsymbol n}$. In turn, the values of $\pi_{a_{\boldsymbol n}}$ and $\pi_{b_{\boldsymbol n}}$ can be obtained from the constraints $H_{\_1}^{\boldsymbol n}$ and $H_{|1}^{\boldsymbol n}$, together with the conditions~\eqref{eq:gauge2}. One gets
\begin{align}
\pi_{a_{\boldsymbol n}} = \frac1{\pi_\alpha}\left(\pi_\varphi\pi_{f_{\boldsymbol n}}+e^{6\alpha}m^2\varphi f_{\boldsymbol n}\right), \qquad
\pi_{b_{\boldsymbol n}} = \frac1{\pi_\alpha}\left[\pi_\varphi\pi_{f_{\boldsymbol n}}-\left(3\pi_\alpha\pi_\varphi-e^{6\alpha}m^2\varphi\right)f_{\boldsymbol n}\right].
\end{align}

After reduction of the system, the Hamiltonian has the same structure as in Eq.~\eqref{eq:H}. The zeroth-order Hamiltonian is still given by Eq.~\eqref{eq:H_0}, while the second-order Hamiltonian can be written in the form~\eqref{eq:H_2} with the coefficients
\begin{subequations}
\begin{align}
E^n_{\pi\pi} &= e^{-2\alpha}\left(1+\frac{3}{n^2-4}\frac{\pi_\varphi^2}{\pi_\alpha^2}\right), \\
E^n_{f\pi} &= -3e^{-2\alpha}\left[\frac{\pi_\varphi^2}{\pi_\alpha}+\frac1{n^2-4}\frac{\pi_\varphi}{\pi_\alpha^2}
\left(3\pi_\alpha\pi_\varphi-e^{6\alpha}m^2\varphi\right)\right], \\
E^n_{ff} &= e^{2\alpha}\left(n^2-1\right)+e^{-2\alpha}\left(9\pi_\varphi^2-6e^{6\alpha}m^2\varphi\frac{\pi_\varphi}{\pi_\alpha}\right)+e^{4\alpha}m^2+\frac{3e^{-2\alpha}}{n^2-4}\frac1{\pi_\alpha^2}\left(3\pi_\alpha\pi_\varphi-e^{6\alpha} m^2\varphi\right)^2.
\end{align}
\end{subequations}

Let us now perform the Mukhanov scaling of the field, completing it into the following canonical transformation (up to the considered perturbative order):
\begin{subequations}
\begin{align}
\bar\alpha = \alpha-\frac{\epsilon^2}2\left(3\frac{\pi_\varphi^2}{\pi_\alpha^2}-1\right)\sum_{\boldsymbol n}f_{\boldsymbol n}^2, \qquad
\bar\pi_{\bar\alpha} = \pi_\alpha+\epsilon^2\sum_{\boldsymbol n}\bigg[-f_{\boldsymbol n}\pi_{f_{\boldsymbol n}}+\bigg(3\frac{\pi_\varphi^2}{\pi_\alpha}+\pi_\alpha\bigg)f_{\boldsymbol n}^2\bigg], \\
\bar\varphi = \varphi+3\epsilon^2\frac{\pi_\varphi}{\pi_\alpha}\sum_{\boldsymbol n} f_{\boldsymbol n}^2, \qquad
\bar f_{\boldsymbol n} = e^\alpha f_{\boldsymbol n}, \qquad
\bar\pi_{\bar{f_{\boldsymbol n}}} = e^{-\alpha}\bigg[\pi_{f_{\boldsymbol n}}-\bigg(3\frac{\pi_\varphi^2}{\pi_\alpha}+\pi_\alpha\bigg)f_{\boldsymbol n}\bigg],
\end{align}
\end{subequations}
whereas $\pi_\varphi$ is left unchanged. This transformation does not alter the structure of the Hamiltonian, but absorbs the dominant contribution in the large-$n$ limit from the terms that couple the matter-field configuration modes with their momenta. The new coefficients of the second-order Hamiltonian are
\begin{subequations}
\begin{align}
&\bar E^n_{\pi\pi} = 1+\frac{3}{n^2-4}\frac{\pi_\varphi^2}{\bar\pi_{\bar\alpha}^2}, \\
&\bar E^n_{f\pi} = \frac{3e^{-2\bar\alpha}}{n^2-4}\,\frac{\pi_\varphi}{\bar\pi_{\bar\alpha}^2}\,\left(3\frac{\pi_\varphi^3}{\bar\pi_{\bar\alpha}}
-2\bar\pi_{\bar\alpha}\pi_\varphi+e^{6\bar\alpha}m^2\bar\varphi\right),
\\
&\bar E^n_{ff} = n^2-\frac12+e^{2\bar\alpha}m^2-\frac32\frac{\pi_\varphi^2}{\bar\pi_{\bar\alpha}^2}-\frac12e^{-4\bar\alpha} \left(\bar\pi_{\bar\alpha}^2-30\pi_\varphi^2\right)-\frac12e^{-4\bar\alpha}\bigg\{27\frac{\pi_\varphi^4}{\bar\pi_{\bar\alpha}^2}
+3e^{6\bar\alpha}m^2\bar\varphi\bigg[8\frac{\pi_\varphi}{\bar\pi_{\bar\alpha}}-\bar\varphi\bigg(3\frac{\pi_\varphi^2}{\bar\pi_{\bar\alpha}^2}-1\bigg)
\bigg]\bigg\}\nonumber\\*
&\quad +\frac3{(n^2-4)}\frac{e^{-4\bar\alpha}}{\bar\pi_{\bar\alpha}^2}\bigg(3\frac{\pi_\varphi^3}{\bar\pi_{\bar\alpha}}-2\bar\pi_{\bar\alpha}\pi_\varphi
+e^{6\bar\alpha}m^2\bar\varphi\bigg)^2.
\end{align}
\end{subequations}

Applying Hamilton's equations in conformal time, one finds that the momentum $\bar\pi_{\bar f_{\boldsymbol n}}$ is still given by Eq.~\eqref{eq:momentum}, except that now
\begin{align}
p_n = -\frac{3\pi_\varphi^2}{(n^2-4)\bar\pi_\alpha^2+3\pi_\varphi^2}, \qquad
q_n = -3e^{-2\bar\alpha}\pi_\varphi\frac{3(\pi_\varphi^3/\bar\pi_{\bar\alpha})-2\bar\pi_{\bar\alpha}\pi_\varphi+e^{6\bar\alpha}m^2\bar\varphi
}{(n^2-4)\bar\pi_\alpha^2+3\pi_\varphi^2}.
\end{align}
As for the equation of motion of $\bar f_{\boldsymbol n}$, Eq.~\eqref{eq:theequation} is still valid as long as we redefine the coefficients in it to be
\begin{subequations}
\begin{align}
r_n &=3 e^{-2\alpha}\frac{\pi_\varphi}{\bar\pi_{\bar\alpha}} \frac{3\pi_\varphi^3-3\bar\pi_{\bar\alpha}^2\pi_\varphi+e^{6\alpha}m^2\bar\varphi(2\bar\pi_{\bar\alpha}-3\bar\varphi\pi_\varphi)
+e^{4\bar\alpha}\pi_\varphi}{(n^2-4)\bar\pi_{\bar\alpha}^2+3\pi_\varphi^2} , \\
s_n &= \frac12+e^{2\bar\alpha}m^2+\frac32\frac{\pi_\varphi^2}{\bar\pi_{\bar\alpha}^2}-\frac12e^{-4\bar\alpha}\bigg(\bar\pi_{\bar\alpha}^2-30\pi_\varphi^2
+27\frac{\pi_\varphi^4}{\bar\pi_{\bar\alpha}^2}\bigg) -\frac12e^{-4\bar\alpha}\bigg[24e^{6\bar\alpha}m^2\bar\varphi\frac{\pi_\varphi}{\bar\pi_{\bar\alpha}}
-3e^{6\bar\alpha}m^2\bar\varphi^2\bigg(3\frac{\pi_\varphi^2}{\bar\pi_{\bar\alpha}^2}-1\bigg)\bigg] +O(n^{-2}).
\end{align}
\end{subequations}
Note that the expressions for $p_n$, $q_n$, $r_n$, and $s_n$ in the new gauge have the same asymptotic behavior as in the gauge adopted in the main body of the article. Owing to this fact, it is not difficult to go through the same arguments and check that the results of Sec. \ref{sec:quantum} apply as well in this case.

The quantities defined via Eqs.~\eqref{eqs:Psi_n1} continue to satisfy a relation of the form $\dot\Psi_{\boldsymbol n}=\Pi_{\Psi_{\boldsymbol n}}$ at leading perturbative order, because Eq.~\eqref{eq:E-v} is expressed in terms of gauge-invariant quantities. Again, they are canonically conjugate, and the equations of motion for $\Psi_{\boldsymbol n}$ are analogous to those for the modes of a Klein-Gordon field with time-dependent mass. Hence, the corresponding CCR's admit a Fock representation which implements the dynamics in a unitary manner. For this, we can introduce two sets of annihilation and creation variables, as in Eqs.~\eqref{eq:creation} and~\eqref{eq:creation2}. The symplectomorphism that relates these variables with the ones adopted in a quantization of the kind discussed in the main text is block diagonal and can be checked to be unitarily implementable, since the coefficients~$\lambda_n$ of its antilinear part still have the form~\eqref{eq:lambda_n}, but with the function~$\chi$ replaced with
\begin{equation}
\bar\chi = \frac{e^{-2\bar\alpha}}{\pi_\varphi}\left(3\frac{\pi_\varphi^3}{\bar\pi_{\bar\alpha}}
-2\bar\pi_{\bar\alpha}\pi_\varphi+e^{6\bar\alpha}m^2\tilde\varphi\right).
\end{equation}
 This new function is well defined if so is the selected gauge (so that, at leading order in the perturbative expansion, the vanishing of $\bar\pi_{\bar\alpha}$ is obviated) and provided, once more, that $\pi_\varphi$ differs from zero.

In conclusion, we see that the results of Sec.~\ref{sec:Bardeen} are reproduced in this gauge, a fact that proves their robustness.


\begin{thebibliography}{99}
\bibitem{note1} In the case of the perturbations, the Poisson bracket between the coefficients of the inhomogeneous modes and their momenta is taken proportional to $1/\epsilon^2$.
\bibitem{lang} The dynamics of this new canonical pair is unitary in this quantization. A time-dependent scaling would spoil the unitarity of the evolution, unless one deals with it before quantizing the system. For instance, a scaling of this type relates the pair with the variables that, in principle, are used in Ref. \cite{Langlois94}.
\end{thebibliography}
\end{document}